\documentclass[twocolumn]{aastex61}
\usepackage{graphicx}
\usepackage{natbib}
\lefthead{The highly obscured DLT16am}
\righthead{Tartaglia et al.}
\shorttitle{The highly obscured DLT16am}
\shortauthors{Tartaglia et al.}

\received{2017-10-07}
\submitjournal{The Astrophysical Journal}

\newcommand{\ha}{H$\alpha$}
\newcommand{\hb}{H$\beta$}

\newcommand{\kms}{\,km\,s$^{-1}$}

\newcommand{\mum}{$\,\mu$m}

\newcommand{\msun}{\,M$_{\sun}$}
\newcommand{\rsun}{\,R$_{\sun}$}

\newcommand{\ang}{\,\AA}

\begin{document}
\title{The early detection and follow-up of the highly obscured Type II supernova 2016ija/DLT16am\footnote{This paper includes data gathered with the 6.5 meter Magellan Telescopes located at Las Campanas Observatory, Chile.}}

\correspondingauthor{L. Tartaglia}
\email{ltartaglia@ucdavis.edu}
\author{L. Tartaglia}
\affiliation{Department of Astronomy and Steward Observatory, University of Arizona, 933 N Cherry Ave, Tucson, AZ 85719, USA}
\affiliation{Department of Physics, University of California, 1 Shields Ave, Davis, CA 95616, USA;}
\author{D.~J. Sand} 
\affiliation{Department of Astronomy and Steward Observatory, University of Arizona, 933 N Cherry Ave, Tucson, AZ 85719, USA}
\author{S.~Valenti} 
\affiliation{Department of Physics, University of California, 1 Shields Ave, Davis, CA 95616, USA;}
\author{S.~Wyatt}
\affiliation{Department of Astronomy and Steward Observatory, University of Arizona, 933 N Cherry Ave, Tucson, AZ 85719, USA}
\author{J.~P.~Anderson}
\affiliation{European Southern Observatory, Alonso de C\'ordova 3107, Casilla 19, Santiago, Chile}
\author{I.~Arcavi}
\affiliation{Department of Physics, University of California, Santa Barbara, CA ,93106-9530, USA}
\affiliation{Las Cumbres Observatory, 6740 Cortona Dr., Suite 102, Goleta, CA 93117, USA}
\affiliation{Einstein Fellow}
\author{C.~Ashall}
\affiliation{Astrophysics Research Institute, Liverpool John Moores University, IC2, Liverpool Science Park, 146 Brownlow Hill, Liverpool L3 5RF, UK}
\author{M.~T.~Botticella}
\affiliation{INAF - Osservatorio Astronomico di Capodimonte, Salita Moiariello 16, Napoli, 80131 Italy}
\author{R.~Cartier}
\affiliation{School of Physics and Astronomy, University of Southampton, Southampton, SO17 1BJ, UK}
\author{T.-W.~Chen}
\affiliation{Max-Planck-Institut f{\"u}r Extraterrestrische Physik, Giessenbachstra\ss e 1, 85748, Garching, Germany}
\author{A.~Cikota}
\affiliation{European Southern Observatory, Karl-Schwarzschild-Str. 2, 85748 Garching b. M\"{u}nchen, Germany}
\author{D.~Coulter}
\affiliation{Department of Astronomy and Astrophysics, University of California, Santa Cruz, CA 95064, USA}
\author{M.~Della Valle}
\affiliation{INAF - Osservatorio Astronomico di Capodimonte, Salita Moiariello 16, Napoli, 80131 Italy}
\author{R.~J.~Foley}
\affiliation{Department of Astronomy and Astrophysics, University of California, Santa Cruz, CA 95064, USA}
\author{A.~Gal-Yam}
\affiliation{Benoziyo Center for Astrophysics, Faculty of Physics, Weizmann Institute of Science, Rehovot 76100, Israel}
\author{L.~Galbany}
\affiliation{PITT PACC, Department of Physics and Astronomy, University of Pittsburgh, Pittsburgh, PA 15260, USA}
\author{C.~Gall}
\affiliation{Dark Cosmology Centre, Niels Bohr Institute, University of Copenhagen, Juliane Maries Vej, 30, 2100 Copenhagen}
\author{J.~B.~Haislip}
\affiliation{Department of Physics and Astronomy, University of North Carolina at Chapel Hill, Chapel Hill, NC 27599}
\author{J.~Harmanen}
\affiliation{Tuorla Observatory, Department of Physics and Astronomy, University of Turku, V\"ais\"al\"antie 20, FI-21500 Piikki\"o, Finland}
\author{G.~Hosseinzadeh} 
\affiliation{Department of Physics, University of California, Santa Barbara, CA ,93106-9530, USA}
\affiliation{Las Cumbres Observatory, 6740 Cortona Dr., Suite 102, Goleta, CA 93117, USA}
\author{D.~A.~Howell}
\affiliation{Department of Physics, University of California, Santa Barbara, CA ,93106-9530, USA}
\affiliation{Las Cumbres Observatory, 6740 Cortona Dr., Suite 102, Goleta, CA 93117, USA}
\author{E.~Y.~Hsiao}
\affiliation{Department of Physics, Florida State University, Keen Building 616, Tallahasee, FL 3206-4350}
\author{C.~Inserra}
\affiliation{School of Physics and Astronomy, University of Southampton, Southampton, SO17 1BJ, UK}
\author{S.~W.~Jha}
\affiliation{Department of Physics and Astronomy, Rutgers, the State University of New Jersey, 136 Frelinghuysen Road, Piscataway, NJ 08854, USA}
\author{E.~Kankare}
\affiliation{Astrophysics Research Centre, School of Mathematics and Physics, Queen's University Belfast, Belfast BT7 1NN, UK}
\author{C.~D.~Kilpatrick}
\affiliation{Department of Astronomy and Astrophysics, University of California, Santa Cruz, CA 95064, USA}
\author{V.~V.~Kouprianov}
\affiliation{Department of Physics and Astronomy, University of North Carolina at Chapel Hill, Chapel Hill, NC 27599}
\author{H.~Kuncarayakti}
\affiliation{Finnish Centre for Astronomy with ESO (FINCA), University of Turku, V\"{a}is\"{a}l\"{a}ntie 20, 21500 Piikki\"{o}, Finland}
\affiliation{Tuorla Observatory, Department of Physics and Astronomy, University of Turku, V\"ais\"al\"antie 20, FI-21500 Piikki\"o, Finland}
\author{T.~J.~Maccarone}
\affiliation{Department of Physics, Texas Tech University, Box 41051, Lubbock, TX 79409-1051, USA}
\author{K.~Maguire}
\affiliation{Astrophysics Research Centre, School of Mathematics and Physics, Queen's University Belfast, Belfast BT7 1NN, UK}
\author{S.~Mattila}
\affiliation{Tuorla Observatory, Department of Physics and Astronomy, University of Turku, V\"ais\"al\"antie 20, FI-21500 Piikki\"o, Finland}
\author{P.~A.~Mazzali}
\affiliation{Astrophysics Research Institute, Liverpool John Moores University, IC2, Liverpool Science Park, 146 Brownlow Hill, Liverpool L3 5RF, UK}
\affiliation{Max-Planck-Institut f\"ur Astrophysik, Karl-Schwarzschild-Str. 1, 85748 Garching bei M\"{u}nchen, Germany}
\author{C.~McCully}
\affiliation{Department of Physics, University of California, Santa Barbara, CA ,93106-9530, USA}
\affiliation{Las Cumbres Observatory, 6740 Cortona Dr., Suite 102, Goleta, CA 93117, USA}
\author{A.~Melandri}
\affiliation{INAF - Osservatorio Astronomico di Brera, via E. Bianchi 36, I-23807 Merate (LC), Italy}
\author{N.~Morrell}
\affiliation{Carnegie Observatories, Las Campanas Observatory, Casilla 601, La Serena, Chile}
\author{M.~M.~Phillips}
\affiliation{Carnegie Observatories, Las Campanas Observatory, Casilla 601, La Serena, Chile}
\author{G.~Pignata}
\affiliation{Departamento de Ciencias Fisicas, Universidad Andres Bello, Avda. Republica 252, Santiago, Chile}
\affiliation{Millennium Institute of Astrophysics (MAS), Nuncio Monse�or S�tero Sanz 100, Providencia, Santiago, Chile}
\author{A.~L.~Piro}
\affiliation{The Observatories of the Carnegie Institution for Science, 813 Santa Barbara St., Pasadena, CA 91101, USA}
\author{S.~Prentice}
\affiliation{Astrophysics Research Institute, Liverpool John Moores University, IC2, Liverpool Science Park, 146 Brownlow Hill, Liverpool L3 5RF, UK}
\author{D.~E.~Reichart} 
\affiliation{Department of Physics and Astronomy, University of North Carolina at Chapel Hill, Chapel Hill, NC 27599}
\author{C.~Rojas-Bravo}
\affiliation{Department of Astronomy and Astrophysics, University of California, Santa Cruz, CA 95064, USA}
\author{S.~J.~Smartt}
\affiliation{Astrophysics Research Centre, School of Mathematics and Physics, Queen's University Belfast, Belfast BT7 1NN, UK}
\author{K.~W.~Smith}
\affiliation{Astrophysics Research Centre, School of Mathematics and Physics, Queen's University Belfast, Belfast BT7 1NN, UK}
\author{J.~Sollerman}
\affiliation{Department of Astronomy and The Oskar Klein Centre, AlbaNova University Center, Stockholm University, SE-106 91 Stockholm, Sweden}
\author{M.~D.~Stritzinger}
\affiliation{Department of Physics and Astronomy, Aarhus University, Ny Munkegade 120, DK-8000 Aarhus C, Denmark}
\author{M.~Sullivan}
\affiliation{School of Physics and Astronomy, University of Southampton, Southampton, SO17 1BJ, UK}
\author{F.~Taddia}
\affiliation{Department of Astronomy and The Oskar Klein Centre, AlbaNova University Center, Stockholm University, SE-106 91 Stockholm, Sweden}
\author{D.~R.~Young}
\affiliation{Astrophysics Research Centre, School of Mathematics and Physics, Queen's University Belfast, Belfast BT7 1NN, UK}

\begin{abstract}
We present our analysis of the Type II supernova DLT16am (SN~2016ija).
The object was discovered during the ongoing $\rm{D}<40\,\rm{Mpc}$ (DLT40) one day cadence supernova search at $r\sim20.1\,\rm{mag}$ in the `edge-on' nearby ($D=20.0\pm1.9\,\rm{Mpc}$) galaxy NGC~1532.
The subsequent prompt and high-cadenced spectroscopic and photometric follow-up revealed a highly extincted transient, with $E(B-V)=1.95\pm0.15\,\rm{mag}$, consistent with a standard extinction law with $R_V=3.1$ and a bright ($M_V=-18.49\pm0.65\,\rm{mag}$) absolute peak-magnitude.
The comparison of the photometric features with those of large samples of Type II supernovae reveals a fast rise for the derived luminosity and a relatively short plateau phase, with a slope of $S_{50V}=0.84\pm0.04\,\rm{mag}/50\,\rm{d}$ consistent with the photometric properties typical of those of fast declining Type II supernovae.
Despite the large uncertainties on the distance and the extinction in the direction of DLT16am, the measured photospheric expansion velocity and the derived absolute $V$-band magnitude at $\sim50\,\rm{d}$ after the explosion match the existing luminosity-velocity relation for Type II supernovae. 
\end{abstract}

\section{Introduction} \label{sec:intro}
The study of cosmic explosions continues to flourish with innovative experiments exploring new regions of discovery space.
Many of these programs are focusing on wide-field imaging and relatively short cadences (e.g. the Palomar Transient Factory - PTF; \citealt{Law09}, the Asteroid Terrestrial-impact Last Alert System - ATLAS; \citealt{ATLAS}, the All Sky Automated Survey for SuperNovae - ASAS-SN; \citealt{Shappee14,Kochanek17}, the PANoramic Survey Telescope And Rapid Response System - Pan-STARRS1; \citealt{panstarrs}, among others), resulting in hundreds of new supernovae (SNe) per year.
This revolution will certainly continue in the era of the Zwicky Transient Facility \citep[ZTF;][]{ZTF}, BlackGEM\footnote{\url{https://astro.ru.nl/blackgem/}} \citep{2015ASPC..496..254B} and the Large Synoptic Survey Telescope \citep[LSST;][]{2008arXiv0805.2366I}.
It is still the case, however, that the nearest SNe are not always caught soon after explosion, relinquishing an opportunity for detailed study of the most observable events -- a recent prominent example was the type Ia SN~2014J in M82 ($D\sim3.5\,\rm{Mpc}$), which was discovered $\sim8\,\rm{days}$ after explosion \citep{Fossey14,2015ApJ...799..106G}.

It is in the hours to days after explosion where clues about the SN progenitors and explosion physics are accessible, and where the fewest observational constraints are available.
Early discovery and prompt multi-wavelength follow-up of nearby SNe is essential to fully characterize the physical properties of stellar explosions.
Except for a few cases where deep archival {\it Hubble Space Telescope (HST)} images are available \citep{smartt09}, one of the best ways to gain insight into a SN progenitor and its explosion mechanism is through the analysis of very early phase data, when the spectra still show the imprint from the outer layers of the progenitor, the explosion energy is still the dominant heat source and the circumstellar medium (CSM) has not yet been overtaken by the SN ejecta.

In Type Ia SNe, the very early light curves and spectra can help to constrain the white dwarf (WD) progenitor radius \citep[e.g.][]{Nugent11,Bloom12,Zheng13}, its $^{56}$Ni distribution \citep[from the early light curve shape; see][]{2014MNRAS.439.1959M,Piro14,2017arXiv170807124M}, to infer the presence of a normal companion star \citep[via direct shocking of the SN Ia ejecta against the normal companion][]{Kasen10,Cao15,Marion16,Hosseinzadeh17} and probe SN Ia explosion mechanisms.
While it is commonly accepted that SNe Ia arise from thermonuclear explosions of carbon-oxygen WDs, it is still unclear by which mechanism(s) the WD accretes the necessary mass \citep[single or double degenerate scenario, see, e.g.,][for a review]{2012PASA...29..447M}.
The detection and strength of \ion{C}{2} (6580\ang) in early spectra may also point to viable explosion mechanisms \citep[see, e.g.,][]{2001MNRAS.321..341M,2002ApJ...568..779H,2007ApJ...668.1132R,2009Natur.460..869K,2010ApJ...725..296F}.

Early light curves of core-collapse (CC) SN shock cooling tails can constrain the progenitor star radius and give useful information about the envelope structure \citep[see, e.g.,][for selected theoretical and observational results]{2011ApJ...728...63R,Bersten12,Arcavi17,Piro17,Sapir17,2017MNRAS.471.2463B}.
Alternatively, `flash spectroscopy' at very early phases, can probe the physical properties of the CSM as well as the mass loss history of the progenitor star prior to its explosion \citep[e.g.][]{Gal-Yam14,Khazov16,2017NatPh..13..510Y}.
Even when archival {\it HST} data are available, radius estimates through the analysis of very early data can give important results, since the progenitor field might be contaminated by the presence of binary companions \citep[e.g.][]{2017ApJ...836L..12T}. 
\begin{figure}
\begin{center}
\epsscale{1.15}
\plotone{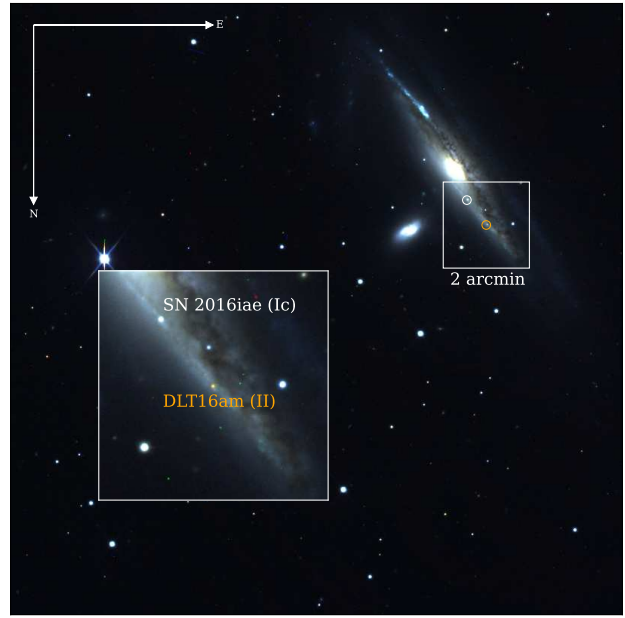}
\caption{RGB images of the DLT16am field obtained using $i$, $r$ and $g$-band images, respectively. The position of the Type Ic SN~2016iae is marked for reference. A zoom-in of the region of the two SNe is shown in the inset ($\sim2$\arcmin). \label{fig:dlt16am}}
\end{center}
\end{figure}

Motivated by the science described above, and by the need for multi-wavelength observations at very early phases, we have begun a pointed, one day cadence SN search for very young transients in the nearby ($D<40\,\rm{Mpc}$; DLT40) Universe.
Given the survey depth of $r\simeq19\,\rm{mag}$ and the proximity of the DLT40 galaxy sample (see Section~\ref{sec:dlt40}), this program is also sensitive to heavily extincted SNe.
A significant number of CCSNe are likely missed by current optical surveys even in normal nearby galaxies due to dust extinction \citep{2012ApJ...756..111M,2017ApJ...837..167J}, which may explain the unexpectedly low CCSNe rate with respect to the cosmic star-formation rate \citep[SFR;][]{2011ApJ...738..154H,2012ApJ...757...70D,2012A&A...545A..96M,2015A&A...584A..62C,2015ApJ...813...93S}.

In this paper we describe the highly obscured, nearby Type II SN DLT16am (SN~2016ija), the first SN discovered by the $D<40\,\rm{Mpc}$ (DLT40) one day cadence supernova search. 
DLT16am was discovered on 2016-11-21.19~UT \citep{2016ATel.9782....1T} in the nearby galaxy NGC~1532 (see Figure~\ref{fig:dlt16am}).
Fortuitously, the type Ic SN~2016iae (Prentice et al., in preparation) was discovered in the same galaxy $\sim2\,\rm{weeks}$ before and was being observed by a number of groups, allowing very tight constraints on the explosion time of DLT16am itself ($\rm{JD}=2457712.6\pm1.0\,\rm{d}$, see Section~\ref{sec:photometry}).
The SN is located at $\rm{RA}$=04:12:07.64, $\rm{Dec}$=$-$32:51:10.57 [J2000], 42\farcs08~E, 76\farcs43~N offset from the center of NGC~1532. 
It was first suggested to be a 91T-like Type Ia SN, showing a nearly featureless and very red continuum, although subsequent early spectra revealed broad \ha~and calcium features on top of a red continuum, leading to a more appropriate classification as a highly reddened Type II SN\footnote{\url{https://wis-tns.weizmann.ac.il/object/2016ija}}.
DLT16am is heavily extincted, but we were still able to obtain a comprehensive multi-wavelength dataset, allowing a detailed comparison of its properties with standard, less extinguished Type II SNe.

This paper is organized as follows:
Section~\ref{sec:dlt40} is a description of the DLT40 survey, while Section~\ref{sec:obs} describes the instrumental setups and the reduction tools used to carry out the follow-up campaign of DLT16am.
In Section~\ref{sec:host} we present the results of our analysis on the host galaxy, NGC~1532.
Sections~\ref{sec:photometry} and \ref{sec:spectroscopy} report the main results of our photometric and spectroscopic analysis, respectively, while in Section~\ref{sec:comparison} we compare the main spectroscopic and photometric features of DLT16am with those displayed by other Type II SNe.
Finally, we summarize our results in Section~\ref{sec:conclusions}.

\section{The DLT40 survey} \label{sec:dlt40}
The goal of DLT40 is not to find many SNe, but $\sim10$ nearby SNe per year within $\sim1\,\rm{day}$ from explosion.
Fully operational since late summer 2016, we observe $\sim300-600$ galaxies per night using a PROMPT $0.4\,\rm{m}$ telescope \citep[PROMPT5;][]{prompt} at the Cerro Tololo Inter-American Observatory (CTIO), achieving a typical single-epoch depth of $r\approx19-20\,\rm{mag}$ with filterless observations and a $45\,\rm{s}$ integration time. 

\begin{figure}
\begin{center}
\epsscale{1.2}
\plotone{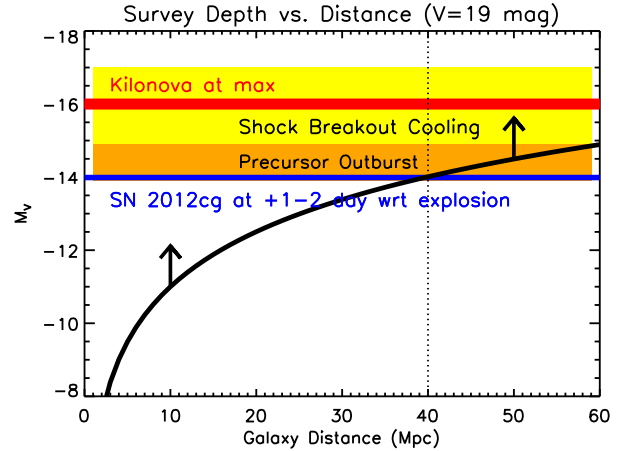}
\caption{Absolute magnitude reached for a survey depth of $V=19\,\rm{mag}$ as a function of distance (black line).  
We highlight the magnitude range for the early SN features necessary to understand their progenitors and explosions. 
Also noted is the absolute magnitude of SN~2012cg $\sim1-2\,\rm{days}$ after explosion, which has possible early time light curve features which can shed light on its progenitor \citep{Marion16} and the $V$-band absolute magnitude reached by the kilonova DLT17ck (AT~2017gfo) at maximum \citep{2017ApJ...848L..24V}. \label{fig:dlt40}}
\end{center}
\end{figure}

The field of view of $10\times10\,\rm{arcmin}^2$ is sufficient to completely image all but the nearest galaxies.
DLT40 is sensitive to most of the early observational signatures that can be used to constrain the nature of the progenitor stars, as illustrated in Figure~\ref{fig:dlt40}.

The DLT40 galaxy sample is drawn from the nearby Gravitational Wave Galaxy Catalogue (GWGC) of \citet{White11}.
From this list, we select all potential DLT40 galaxies with $M_{B}<-18\,\rm{mag}$, Milky Way (MW) extinction $A_{V}<0.5\,\rm{mag}$ and recessional velocity $v_r<3000$\kms, corresponding to a Hubble flow distance of $D\lesssim40\,\rm{Mpc}$.
Our nightly schedule is based on this list, with preference given to those fields observed in the previous three days, along with intrinsically bright ($M_{B}<-20\,\rm{mag}$) or nearby galaxies ($D<11\,\rm{Mpc}$) and fields that have more than one target galaxy within the PROMPT field of view.
Given the high star formation rates and stellar masses of our targets, and based on current SN rates and simulations, we expect to find $\sim20\,\rm{SN}\,\rm{yr^{-1}}$. 
Assuming some weather and instrument problems, we ultimately expect $\sim10\,\rm{SN}\,\rm{yr^{-1}}$ ($\sim2-3$ Ia, $\sim2-3$ Ib/c and $\sim4-5$ Type II SNe per year) discovered within $\sim1\,\rm{d}$ of explosion. 
PROMPT5 DLT40 pre-reduced images are delivered within $\sim1\,\rm{minute}$ of the data being taken, and are further processed by our pipeline, which includes quality checks, difference imaging \citep[with {\sc HOTPANTS};][]{Becker15}, candidate detection and scoring (based on the difference image properties), databasing and web site generation of stamps for candidate inspection. 
The average lag between an observation and the visualization of the SN candidates is $\sim3-4\,\rm{minutes}$, after which one or more team members can immediately trigger a PROMPT5 confirmation image or further observations at other facilities.

First results from the DLT40 survey were presented for the nearby Type Ia SN~2017cbv (DLT17u) in \citet{Hosseinzadeh17}, and 13 discoveries have been reported to the Astronomer's Telegram service \citep{DLT16w,DLT16am,DLT17h,DLT17u,DLT17ah,DLT17ar,DLT17aw,DLT17bk,DLT17bx,DLT17cd,DLT17cg,DLT17ch,DLT17cq}, in line with ​our initial expectations.
More DLT40 results are in preparation.

\section{Observations And Data Reduction} \label{sec:obs}
\begin{center}
\startlongtable
\begin{deluxetable*}{ccccccc}
\tablecaption{Log of the spectroscopical observations. \label{table:speclog}}
\tablehead{\colhead{Date (UT)} & \colhead{JD} & \colhead{Phase} & \colhead{Instrumental setup} & \colhead{Grism / Grating} & \colhead{Spectral range} & \colhead{Exposure time} \\ 
\colhead{} & \colhead{} & \colhead{(d)} & \colhead{} & \colhead{} & \colhead{(\AA)} & \colhead{(s)} } 
\startdata
20161122 & 2457715.144 & $+2$   & FTS+FLOYDS       & $235\,\rm{l/mm}$ & $5000-9200$  & 3600 \\
20161123 & 2457716.557 & $+4$   & NOT+ALFOSC   & Gr4              & $4000-9700$  & 2400 \\
20161124 & 2457717.315 & $+5$   & SALT+RSS     & PG0300           & $3300-9200$  & 2326 \\
20161127 & 2457719.735 & $+7$   & ESO NTT+SOFI     & GB+GR            & $9000-22000$ & $8\times270+8\times450$ \\
20161201 & 2457723.753 & $+11$  & ESO NTT+EFOSC    & Gr16             & $6000-10000$ & 2700 \\
20161213 & 2457736.516 & $+24$  & NOT+ALFOSC   & Gr4              & $4000-9700$  & 2400 \\
20161219 & 2457742.732 & $+30$  & ESO NTT+EFOSC    & Gr16             & $6000-10000$ & 2700 \\
20161222 & 2457744.517 & $+32$ & ESO VLT+X-shooter & UVB+VIS+NIR & $3500-25000$ & 700+600+600 \\
20161222 & 2457744.881 & $+32$  & FTN+FLOYDS       & $235\,\rm{l/mm}$ & $5000-9200$  & 3600 \\
20161224 & 2457746.568 & $+34$  & Gemini+GNIRS & ShortXD        & $8800-24000$ & 3000 \\
20161225 & 2457747.846 & $+35$  & FTN+FLOYDS       & $235\,\rm{l/mm}$ & $5000-9200$  & 3600 \\
20161227 & 2457749.842 & $+37$  & FTN+FLOYDS       & $235\,\rm{l/mm}$ & $5000-9200$  & 3600 \\
20170102 & 2457755.671 & $+43$  & Baade+FIRE   & LDPrism          & $8800-20000$ & $8\times126.8$    \\  
20170102 & 2457756.019 & $+43$  & FTS+FLOYDS       & $235\,\rm{l/mm}$ & $5000-9200$  & 3600 \\
20170104 & 2457758.693 & $+46$  & ESO NTT+SOFI     & GB+GR            & $9000-24000$ & $8\times270+8\times450$ \\        
20170105 & 2457758.945 & $+46$  & FTS+FLOYDS       & $235\,\rm{l/mm}$ & $5000-9200$  & 3600 \\
20170107 & 2457760.928 & $+48$  & FTS+FLOYDS       & $235\,\rm{l/mm}$ & $5000-9200$  & 3600 \\
20170113 & 2457766.815 & $+54$  & Gemini+GNIRS & ShortXD          & $8800-24000$ & 3000 \\
20170116 & 2457770.386 & $+58$  & NOT+ALFOSC   & Gr4              & $4000-9700$  & 2400 \\
20170117 & 2457771.620 & $+59$  & ESO NTT+EFOSC    & Gr16             & $6000-10000$ & 2400 \\
20170118 & 2457772.620 & $+60$  & ESO NTT+SOFI     & GB+GR            & $9000-24000$ & $8\times270+8\times450$ \\        
20170125 & 2457779.635 & $+67$  & ESO NTT+EFOSC    & Gr16             & $6000-10000$ & 2400 \\
20170204 & 2457789.619 & $+77$  & ESO NTT+EFOSC    & Gr16             & $6000-10000$ & 2700 \\
20170206 & 2457790.591 & $+78$  & ESO NTT+SOFI     & GB+GR            & $9000-24000$ & $6\times270+6\times450$ \\
20170219 & 2457803.559 & $+91$  & ESO NTT+EFOSC    & Gr16             & $6000-10000$ & 2700 \\
20170224 & 2457809.554 & $+97$  & ESO NTT+EFOSC    & Gr16             & $6000-10000$ & 2700 \\
20170307 & 2457819.580 & $+107$ & ESO NTT+SOFI     & GB               & $9000-17000$ & $6\times270$ \\
20170307 & 2457820.510 & $+108$ & ESO NTT+EFOSC    & Gr16             & $6000-10000$ & $2700$ \\
\enddata
\tablecomments{FTN: $2\,\rm{m}$ Faulkes Telescope North, Las Cumbres Observatory node at the Haleakala Observatory, Hawaii; FTS: $2\,\rm{m}$ Faulkes Telescope South, Las Cumbres Observatory node at the Siding Spring Observatory, Australia; NOT: $2.56\,\rm{m}$ Nordic Optical Telescope, located at Roque de los Muchachos, La Palma, Spain; ESO VLT: $8.2\,\rm{m}$ Very Large Telescope, located at the ESO Cerro Paranal Observatory, Chile; ESO NTT: $3.58\,\rm{m}$ New Technology Telescope, located at the ESO La Silla Observatory, Chile; Baade: $6.5\,\rm{m}$ Magellan (Walter Baade) Telescope at the Las Campanas Observatory, Chile; Gemini: $8.19\,\rm{m}$ Gemini North Telescope at the Mauna Kea Observatories, Hawaii.}
\end{deluxetable*}
\end{center}
\begin{figure*}
\begin{center}
\epsscale{1.15}
\plotone{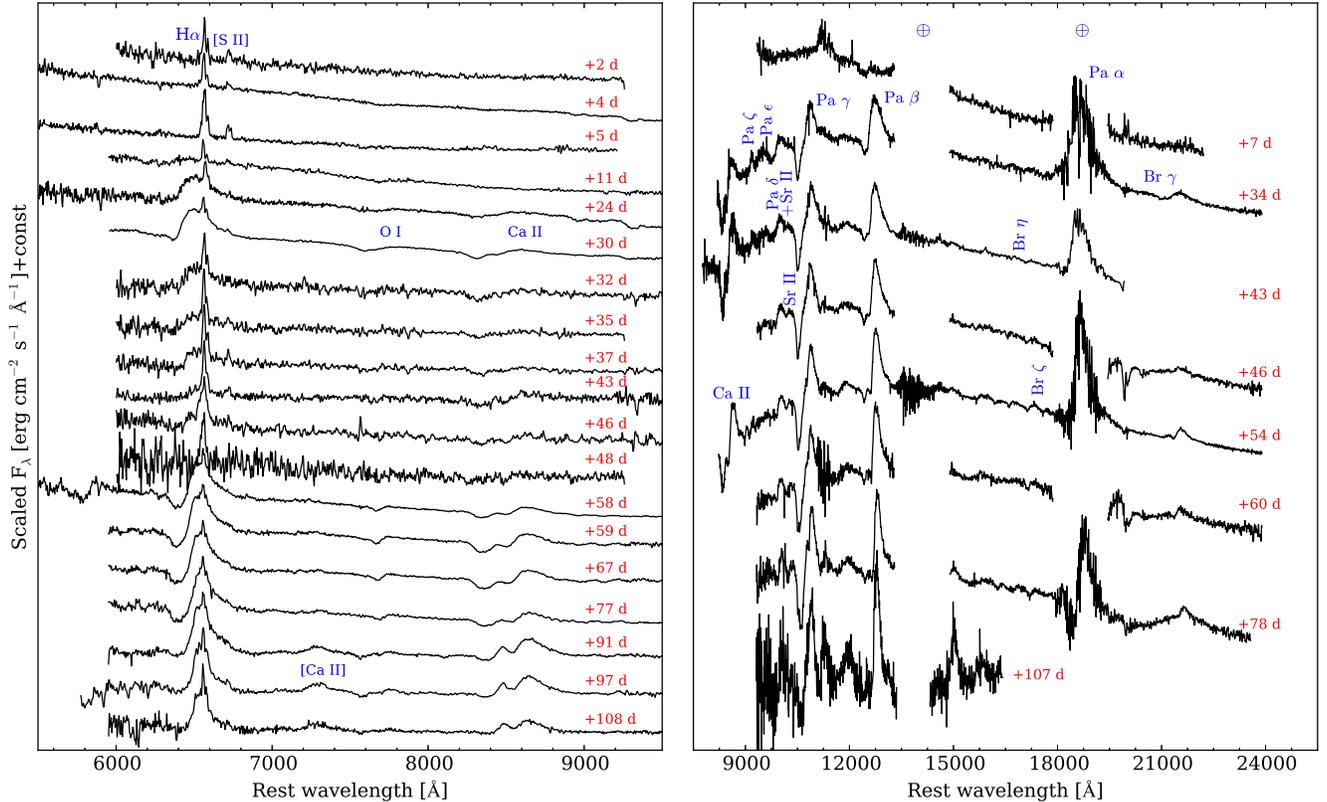}
\caption{{\bf Left:} Optical and {\bf Right:} NIR spectral sequence of DLT16am. Phases refer to the epoch of the explosion. $\oplus$ symbols mark the positions of the main absorption telluric features.
Optical spectra have been corrected for total extinction and redshift. The positions of the main lines are marked. NIR spectra have not been corrected for the total extinction to highlight shallower spectroscopic features. \label{fig:spectra}}
\end{center}
\end{figure*}

\subsection{Spectra} \label{sec:spectra}
The spectral sequence of DLT16am is shown in Figure~\ref{fig:spectra}, while the log of the spectroscopic observations in the optical and NIR domains is reported in Table~\ref{table:speclog}.
Optical spectra were mainly provided by PESSTO using the ESO $3.58\,\rm{m}$ NTT with the ESO Faint Object Spectrograph and Camera \citep[v.2, EFOSC2][]{1984Msngr..38....9B} and the Las Cumbres Observatory network of telescopes, using the $2\,\rm{m}$ Faulkes North and South telescopes with FLOYDS and reduced as in \cite{valenti14}.
Early phase spectra were also provided using the $10\,\rm{m}$ South African Large Telescope (SALT) with the Robert Stobie Spectrograph (RSS) and the $2.56\,\rm{m}$ NOT with ALFOSC.
NOT spectra were reduced using \textsc{Foscgui}, while the SALT spectrum was reduced using a dedicated pipeline \citep[\textsc{PySALT;}][]{pysalt}.

NIR spectra were obtained using the standard `ABBA' technique and an A0V telluric standard was observed at similar airmass in order to simultaneously correct for telluric absorption and to flux calibrate the main science data, and were mainly provided by PESSTO using NTT with SOFI \citep[all reduced as in ][]{smartt15}.
Two NIR spectra were obtained using the Gemini NIR Spectrograph (GNIRS) at Gemini North \citep{GNIRS} in cross-dispersed mode, using the $32\,\rm{l}\,\rm{mm^{-1}}$ grating and the 0\farcs675 slit.
This setup yields a continuous wavelength coverage from 0.8 to 2.5\mum~with a resolution of $R\sim1000$.  The data were taken with the slit along the parallactic angle, and were reduced with the {\sc XDGNIRS} PyRAF-based pipeline provided by Gemini Observatory. 
Flux calibration and telluric correction were performed following the methodology of \citet{Vacca03}.
A NIR spectrum was also obtained using the $6.5\,\rm{m}$ Magellan Baade Telescope with the Folded-port InfraRed Echellette (FIRE). 
The spectrum was taken in the long slit prism mode, 0\farcs6~slit width, and was a combination of 8 exposures of $126.8\,\rm{s}$ each.
The reduction was done using the standard `firehose' IDL package \citep{2013PASP..125..270S}.

Multi-wavelength ($300-2500\,\rm{nm}$) intermediate resolution spectra were obtained using the ESO {\it Very Large Telescope} (VLT) with X-shooter \citep{2011A&A...536A.105V}, mounted at the Cassegrain focus of the $8\,\rm{m}$ VLT UT2 telescope.
UVB, VIS and NIR arm data (covering the 300--559.5, 559.5--1024 and 1024--2480$\,\rm{nm}$ wavelength ranges respectively) were reduced using the X-shooter dedicated pipeline through the \textsc{esoreflex} environment \citep{esoreflex}.

Optical and NIR spectra will be released through the Weizmann Interactive Supernova data REPository \citep[WISEREP\footnote{\url{https://wiserep.weizmann.ac.il/}};][]{2012PASP..124..668Y}.

\subsection{Light curves} \label{sec:lightcurves}
\begin{figure}
\begin{center}
\epsscale{1.15}
\plotone{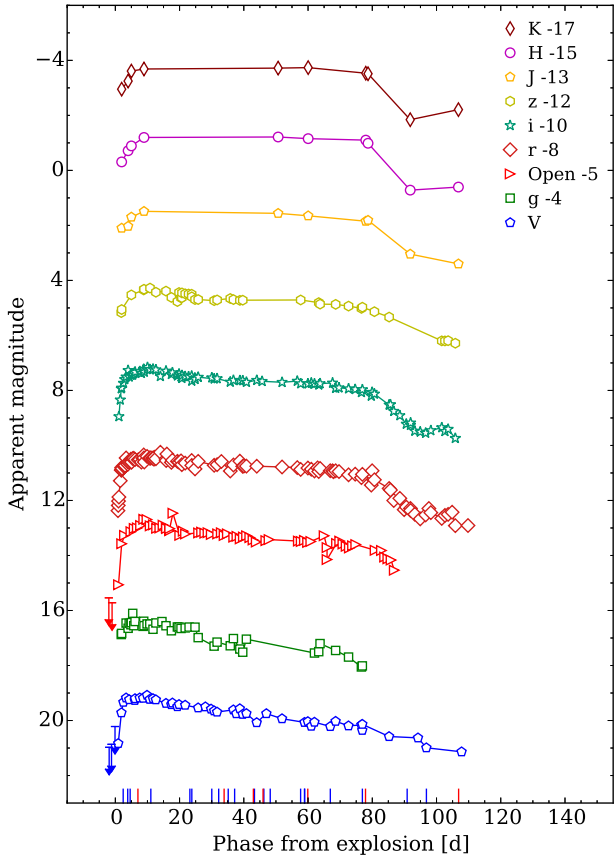}
\caption{Multi band light curves of DLT16am from the optical to the NIR. Blue ticks mark the epochs at which optical spectra were obtained, while red ticks those of the NIR spectra. $BVJHK$ and $griz$ magnitudes were calibrated to the Vega and to the AB photometric system, respectively. Phases refer to the estimated epoch of the explosion (see Sections~\ref{sec:intro} and \ref{sec:photometry} for details). Magnitudes were not corrected for MW or host extinction. Main pre-discovery limits are reported for $V$-band and unfiltered light curves. \label{fig:phot}}
\end{center}
\end{figure}

Photometric data are shown in Figure~\ref{fig:phot} and reported in Tables~\ref{table:BVphot}, \ref{table:grizphot} and \ref{table:JHKphot}, including publicly available early time photometry \citep{2016ATel.9784....1S,2016ATel.9789....1C}.

The {\it griz} data were mainly provided by the Las Cumbres Observatory \citep{Brown13}.
Additional {\it griz} data were obtained using the MPG/ESO $2.2\,\rm{m}$ telescope at the La Silla Observatory with the Gamma-ray Burst Optical/Near-infrared Detector \citep[GROND;][]{2008PASP..120..405G}, which also provided early-phase photometric data at near infrared (NIR) wavelengths, the $2\,\rm{m}$ Liverpool Telescope (LT) with the optical imaging component of the Infrared-Optical camera (IO:O), the $2.56\,\rm{m}$ Nordic Optical Telescope (NOT) with the Andalucia Faint Object Spectrograph and Camera (ALFOSC) and Las Campanas Observatory $1\,\rm{m}$ Swope Telescope with an E2V camera \citep{1973ApOpt..12.1430B}. 
Pre-reduction steps (including bias, flat-field corrections, image stacking and astrometry calibration) for GROND frames were performed as in \cite{2008ApJ...685..376K}, while the final magnitudes for GROND, IO:O, ALFOSC and Swope frames were obtained using the reduction pipeline \textsc{SNOoPy\footnote{\url{http://sngroup.oapd.inaf.it/snoopy.html}}}.
All these images were template subtracted and photometry was calibrated to the AAVSO Photometric All-Sky Survey (APASS\footnote{\url{https://www.aavso.org/apass}}) catalog.

DLT16am was also observed in the framework of the CHilean Automatic Supernova sEarch (CHASE) survey \citep{2009AIPC.1111..551P} in Johnson-Cousins $R$ filters using the PROMPT1 telescope \citep{prompt}.
All images were reduced following standard procedures, including dark (with the same exposure time) and flat-field corrections and then template subtracted. 
The computed photometry is relative to a local sequence calibrated in the field of NGC~1532.
The PROMPT1 Johnson-Cousins $R$-band magnitudes were transformed to the Sloan $r$-band following the procedure reported in \cite{2008MNRAS.388..971P}.

Unfiltered data were provided by the DLT40 (see Section~\ref{sec:dlt40}) SN search using the $0.4\,\rm{m}$ PROMPT5 telescope \citep{prompt}, template subtracted and calibrated to the APASS $r$-band.

DLT16am exploded in the same host and $\sim2\,\rm{weeks}$ after the Type Ic SN~2016iae \citep{2016ATel.9754....1J} and pre-SN $V$-band acquisition images of the field were obtained during its PESSTO spectroscopic campaign.
These data were used to constrain the explosion epoch and are included in the $V$-band light curve of DLT16am (see \citealt{2016ATel.9784....1S} and Section~\ref{sec:photometry}). \\

{\it JHK} data were also provided by the Public ESO Spectroscopic Survey for Transient Objects (PESSTO\footnote{\url{http://www.pessto.org}}) using the $3.58\,\rm{m}$ New Technology Telescope (NTT) with the Son Of ISAAC camera \citep[SOFI;][]{1998Msngr..91....9M} and reduced using their dedicated pipeline \citep[see][]{smartt15} and the NOT Unbiased Transient Survey (NUTS\footnote{\url{http://csp2.lco.cl/not/}}) with the NOT near-infrared Camera and spectrograph (NOTCam).
Pre-reduction steps for NOTCam data were obtained with a modified version of the external \textsc{iraf} package \textsc{notcam} (v.2.5\footnote{http://www.not.iac.es/instruments/notcam/guide/observe.html\\\#reductions}).
In addition to differential flat-fielding and corrected using the median of the sky level at the time of the observations, a bad pixel masking and distortion correction were applied before stacking dithered images.
Magnitudes were obtained from pre-reduced images using a dedicated pipeline (\textsc{Foscgui}\footnote{\url{http://sngroup.oapd.inaf.it/foscgui.html}}).
NIR photometry was calibrated to the Two Micron All-Sky Survey (2MASS\footnote{\url{https://www.ipac.caltech.edu/2mass/}}) catalog, through point-spread-function (PSF) fitting techniques.

DLT16am was also observed with $\textit{Swift}$/XRT on 2016 November 25 (for $2889.4\,\rm{s}$), November 26 (for $2966.8\,\rm{s}$), November 28 (for $1023.9\,\rm{s}$) and December 1 (for $2936.8\,\rm{s}$).
A previous $\textit{Swift}$/XRT exposure of SN~2016iae (which is in the same galaxy) was used to extract the background in the region of DLT16am.
Due to this complicated background, we obtained a limiting count rate (assuming an 18\arcsec~radius) of $1.67\times10^{-3}\,\rm{counts}\,\rm{s^{-1}}$. 
Assuming a power-law model with a photon index of 2 and a Galactic absorption $1.58\times10^{20}\,\rm{cm}^{-2}$ \citep{2005A&A...440..775K}, this corresponds to an unabsorbed flux $<4.1\times10^{-14}\,\rm{erg}\,\rm{cm^{-2}}\,\rm{s^{-1}}$ ($0.4-5\,\rm{keV}$) and an approximate luminosity of $<1.9\times10^{39}\,\rm{erg}\,\rm{s^{-1}}$ at $20\,\rm{Mpc}$ (see Section~\ref{sec:host}).
We could not detect DLT16am in UVOT frames, due to the high extinction in the direction of DLT16am (see Section~\ref{sec:reddening}).

\section{The host galaxy} \label{sec:host}
NGC~1532, the host of DLT16am, is a SB(s)b edge-on galaxy \citep{1991rc3..book.....D}, located at RA$=$04:12:04.3, Dec$=-$32:52:27~[J2000], with an apparent total magnitude of $B=10.65\pm0.09\,\rm{mag}$ \citep{trove.nla.gov.au/work/8318286}, showing prominent dust lanes close to the position of DLT16am. 
From our X-shooter spectrum obtained on 2016 December 22, we infer an heliocentric velocity of 1367\kms ($\rm{z}=0.00456$, see Section~\ref{sec:opt_spectra}), estimated from the average positions of the Balmer emission lines.
Throughout the paper we will adopt a distance of $20.0\pm1.8\,\rm{Mpc}$ \citep{2013AJ....146...86T} to NGC~1532 (leading to a projected distance of $\simeq8.4\,\rm{kpc}$ from the center of the host for DLT16am), although a wide range of values are reported in the literature, suggesting a larger uncertainty.
In Section~\ref{sec:comparison}, we will show that this value, along with the derived extinction (see Section~\ref{sec:reddening}), gives absolute magnitudes, pseudo-bolometric luminosities, and hence an estimated $^{56}\rm{Ni}$ mass consistent with other distance and reddening-independent quantities.
\begin{figure}
\begin{center}
\epsscale{1.15}
\plotone{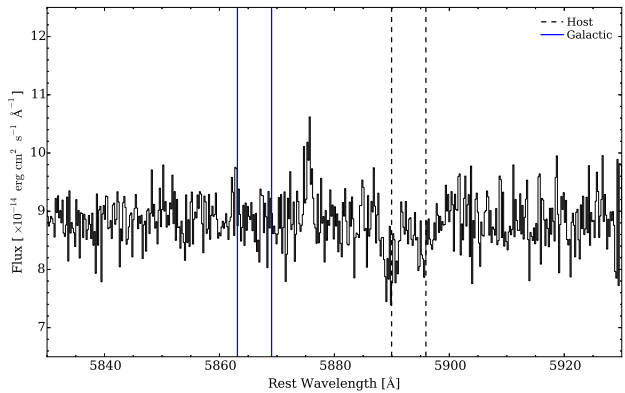}
\caption{Zoom-in of the region of the \ion{Na}{1D} doublet in our X-shooter spectrum. The positions of the host and Galactic \ion{Na}{1D} lines are marked with black dashed and blue solid lines, respectively. \label{fig:sodium}}
\end{center}
\end{figure}

\subsection{Extinction estimate} \label{sec:reddening}
While for the foreground Galactic extinction we assumed $A_V=0.042\,\rm{mag}$, as derived from the \citet{2011ApJ...737..103S} IR-based dust map, the determination of the host reddening in the direction of DLT16am proved to be more complicated. 
Early spectra of DLT16am exhibit a very red continuum, with low or almost no signal at wavelengths $<6000$\ang~(with the exception of very early phases, $T\lesssim5\,\rm{d}$, see Section~\ref{sec:opt_spectra}), a signature of very high reddening in the direction of the transient.
\begin{figure*}
\begin{center}
\epsscale{1.15}
\plotone{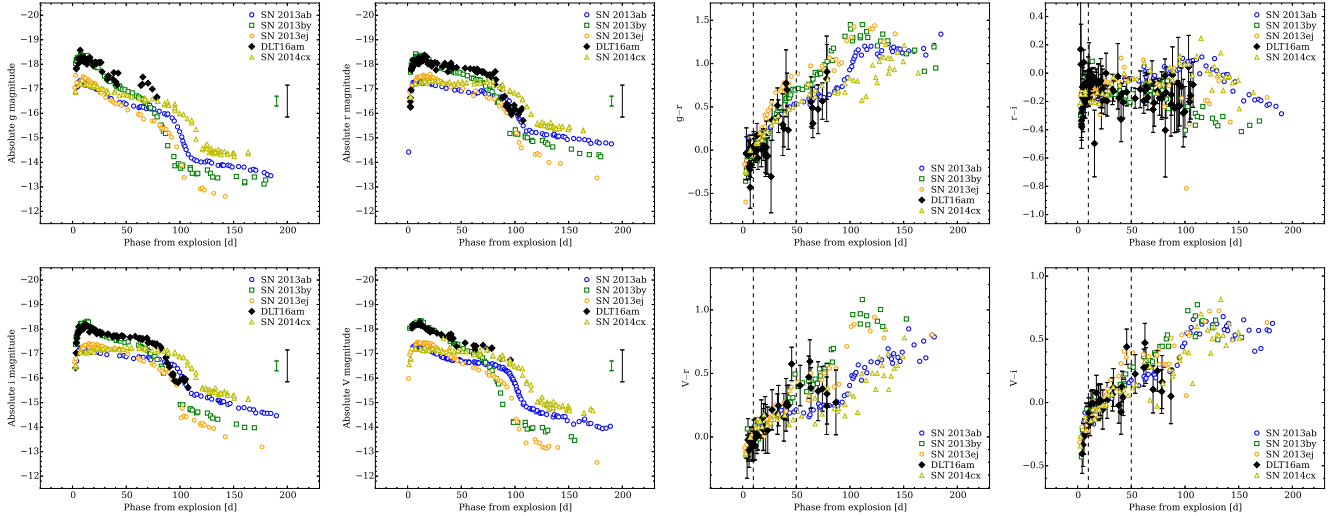}
\caption{Absolute $g$, $V$, $r$ and $i$-band light curves ({\bf left}) and  $g-r$, $r-i$, $V-r$, $V-i$ color curves ({\bf right}) of DLT16am adopting a contribution of $E(B-V)=1.95\pm0.15\,\rm{mag}$ from the host galaxy to the total extinction.
Absolute light and color curves of the models adopted to infer the host extinction (see the text for more details) are also shown for comparison.
The representative error bar at $+200\,\rm{d}$ and $-16\,\rm{mag}$ in the absolute curve panels corresponds to the uncertainty on the derived extinction in the direction of DLT16am, while the greend errorbar corresponds to the uncertainty on the distance modulus.
Dashed black vertical lines in the color curves delimit the region where the colors of DLT16am were fitted to the references (see the text for additional details). \label{fig:abs_colors}}
\end{center}
\end{figure*}

In order to get an estimate of the host galaxy reddening, we measured the equivalent widths (EWs) of the \ion{Na}{1D} doublet in our X-shooter spectrum (see Figure~\ref{fig:sodium}), on the basis of the correlation between the strength of the \ion{Na}{1D} $\lambda\lambda5890$,5896 absorption and the color excess \citep[see][]{2012MNRAS.426.1465P,2013ApJ...779...38P}.

The estimated values ($\rm{EW_{D1}\simeq1.7}$\ang, $\rm{EW_{D2}\simeq2.5}$\ang) are significantly larger than the saturation limit, where the relation flattens \citep[$\rm{EW}\gtrsim0.2$\ang; see][]{2012MNRAS.426.1465P}.
We therefore estimated the reddening by matching the colors of DLT16am, during the plateau phase, with those of other Type II SNe \citep[i.e. SNe~2013ej, 2013ab, 2013by and 2014cx;][respectively, see Figure~\ref{fig:abs_colors}]{valenti14,2015MNRAS.450.2373B,2015MNRAS.448.2608V,2016ApJ...832..139H}.
These transients were selected among those with similar plateau lengths and observed using similar filters and, for the same reasons, used throughout the paper as comparison objects.

During the plateau phase, the outer hydrogen layer starts to recombine as its temperature decreases to $\sim6000\,\rm{K}$, until the recombination front reaches its base and the plateau ends.
This recombination temperature of hydrogen is relatively insensitive to density and metallicity and hence during the plateau SNe II typically share similar physical conditions \citep[see, e.g.,][]{1992ApJ...395..366S}.
A scatter in their colors may therefore be attributed to extinction rather than other intrinsic behaviors \citep[see, e.g.,][and references therein]{2014MNRAS.442..844F}.
We therefore fitted the color excess $E(B-V)$ to match the available colors ($g-r$, $r-i$, $V-r$ and $V-i$) of the comparison objects (hereafter `references') within the plateau phase, after correcting the observed colors of the references for the corresponding total extinction. 
A set of values were obtained comparing the colors of DLT16am to those of each reference, taking the minima of the $\chi^2$ distributions obtained for each color (i.e. $\chi^2=\sum_{k}\frac{1}{\sigma^2_k}(col_{k,ref}-col_{k,\rm{DLT16am}})^2$, where $\sigma^2_k$ are the errors on the colors of DLT16am, and $k=g-r,\,r-i,\,V-r,\,V-i$).

Different extinction laws were recently proposed after analyzing heavily extincted objects \citep[see, e.g., the case of the obscured Type Ia SN~2002cv][]{2008MNRAS.384..107E}.
Following \citet{2000ApJ...533..682C} we also tested their proposed extinction law with $R_V=4.05\pm0.80$, getting a reasonable fit to the colors of the references, but unusual bright absolute peak magnitudes for DLT16am, although still within the combined errors on the reddening and distance.
For this reason, we cannot rule out a different extinction law for the environment of DLT16am, although we remark that changing $R_V$ did not significantly improve the result of the fit.
Hereafter, we will therefore adopt a standard value \citep[$R_V=3.1$;][]{1989ApJ...345..245C}.
Averaging the best fit values obtained using the different references we find $E(B-V)=1.95\pm0.15\,\rm{mag}$.
A comparable value was obtained measuring the Balmer decrement (i.e. $E(B-V)=2.0\,\rm{mag}$ through the \ha/\hb~flux ratio) from our X-shooter spectrum, assuming a case B recombination ratio and a standard extinction law with $R_V=3.1$ \citep[see, e.g.,][]{2012A&A...537A.132B}.

In Figure~\ref{fig:abs_colors} we show the resulting color evolution and absolute magnitudes of DLT16am compared with those of the references.
We find color evolutions comparable with those of the models, while DLT16am shows brighter absolute magnitudes than those displayed by other objects (although similar to those observed in SN~2013by).
In Section~\ref{sec:comparison} we will show that the derived absolute magnitude is consistent with the photospheric expansion velocity derived from the spectroscopic analysis, matching the existing luminosity-velocity relation for SNe II.

\subsection{Metallicity and star formation rate} \label{sec:host_physics}
\begin{figure}
\begin{center}
\epsscale{1.1}
\plotone{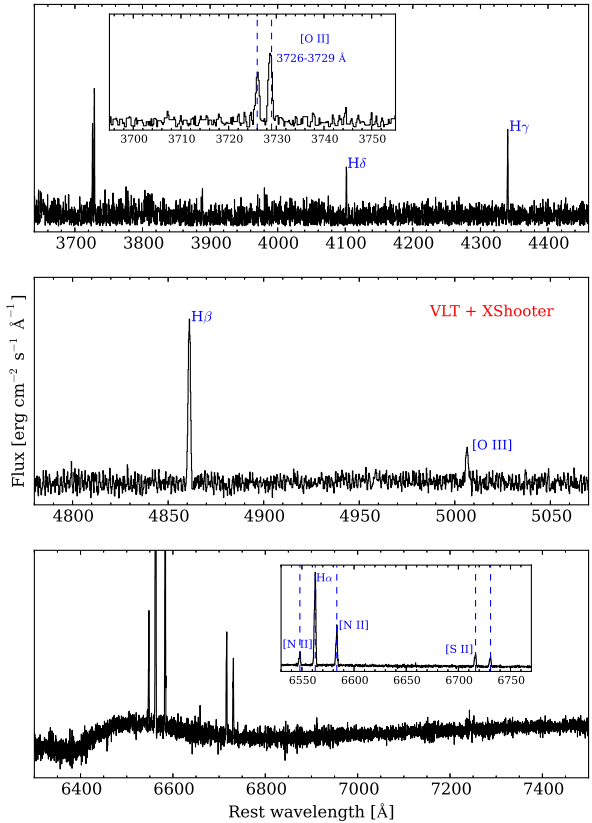}
\caption{UVB and VIS X-shooter spectra obtained on 2016 December 22. Narrow lines emitted by the host galaxy are identified. Insets in top and bottom panels show zoom-ins around the main spectral features. \label{fig:xshoo_spectra}}
\end{center}
\end{figure}

After correcting our X-shooter UVB and VIS spectra for the foreground Galactic extinction, redshift (using the values reported above) and host galaxy extinction, we estimated the local metallicity and star formation rate (SFR) of NGC~1532 at the position of DLT16am.
An identification of the host galaxy lines commonly used in the literature is reported in Figure~\ref{fig:xshoo_spectra}.

Using the calibration of \citet{2011MNRAS.412.1145P}, based on the strong emission lines of $\rm{O}^{++}$, $\rm{N}^{+}$ and $\rm{S}^{+}$ (the NS calibration), we estimate a local metallicity of $12+\log{(O/H)}=8.45\,\rm{dex}$ or $12+\log{(N/H)}=7.46\,\rm{dex}$, while following \citet{2004MNRAS.348L..59P}, we obtain $12+\log{(O/H)}=8.67\,\rm{dex}$ and $8.84\,\rm{dex}$ using the $N2$ \citep{2002RMxAC..12..257D} and their redefinition of the $O3N2$ \citep{alloin79} indices, respectively.
Assuming a solar value of $12+\log{(O/H)}=8.69\,\rm{dex}$ \citep{2009ARA&A..47..481A}, these correspond to $\sim\rm{Z}_{\odot}$, which is larger than the mean values found by \citet{2016A&A...589A.110A} for a sample of Type II SNe, and might be even higher, since metallicity values estimated through line diagnostic are believed to underestimate local abundances \citep[see, e.g.,][]{2012MNRAS.426.2630L}.

Following \citet{2002MNRAS.332..283R}, we derive the local SFR from the luminosities of \ha, using the relation given by \citet{1994ApJ...435...22K}:
\begin{equation} \label{eq:sfr_halpha}
SFR(\mathrm{H\alpha})(\mathrm{M_\odot}\,\mathrm{yr^{-1}})=7.9\times10^{-42} L_{\mathrm{H\alpha}}\,(\rm{erg}\,\rm{s^{-1}}).
\end{equation}
Accounting for the derived distance of NGC~1532, we obtain a local SFR of $1.64\times10^{-1}\,\rm{M_{\odot}}\,\rm{yr^{-1}}$.

\section{Photometry} \label{sec:photometry}
Roughly $2\,\rm{weeks}$ before the discovery of DLT16am, the Type Ic SN 2016iae exploded in NGC~1532 \citep{2016ATel.9749....1T}.
Pre-SN images of DLT16am were collected during the photometric follow-up campaign of SN~2016iae carried out with the Las Cumbres Observatory $1\,\rm{m}$ telescope network and these provided the template images used in our photometric analysis. 
GROND data and NTT acquisition images of SN~2016iae were used as templates for frames obtained with the same instrumental setup.

Prior to the discovery on 2016-11-21.19~UT \citep{2016ATel.9782....1T}, the last DLT40 non-detection was on 2016-11-19.19~UT ($R>20.7\,\rm{mag}$), suggesting $\rm{JD}=2457712.7\pm1.0\,\rm{d}$ as the explosion epoch for DLT16am.
On the other hand, after the initial discovery, \citep{2016ATel.9784....1S} reported previous detections of the transient on PESSTO $V$-band acquisition images of SN~2016iae.
Although they report a marginal detection on 2016-11-20.10~UT ($\rm{JD}=2457712.6$) analyzing these archival images we could not find any point source within $\sim3$\arcsec~from the position of DLT16am, while we clearly detected the transient on 2016-11-21.10~UT ($\rm{JD}=2457713.6$).
Since the detection limit in the frame obtained on November 20.10~UT is not deep enough to rule out the presence of the transient at this time ($V>20.2\,\rm{mag}$), we will assume $2457712.6\pm1.0\,\rm{d}$ as explosion epoch, and refer to this date in computing the phases in both our photometric and spectroscopic analysis.

The final apparent light curves are shown in Figure~\ref{fig:phot}, while Figure~\ref{fig:abs_colors} shows the absolute and color curves obtained after correcting for the host galaxy extinction (see Section~\ref{sec:reddening}).
Due to the high extinction [$E(B-V)=1.95\pm0.15\,\rm{mag}$ see Section~\ref{sec:reddening}], DLT4016am was never detected in the $B$ band during the first $\sim60\,\rm{d}$ after the explosion, while early detections covering the rise to the light curve maximum at redder wavelengths were available.
At early phases, the light curves show relatively fast rises to the maximum in all bands, after which they set on a plateau-like phase with roughly constant magnitudes up to $\simeq+80\,\rm{d}$.

The early evolution ($\lesssim10\,\rm{d}$) of the $g-r$, $r-i$, $V-r$ and $V-i$ colors is fast, suggesting a rapid temperature evolution (see also Section~\ref{sec:spectroscopy}), supporting our claim that DLT16am was discovered soon after the explosion.

\begin{figure}
\begin{center}
\epsscale{1.15}
\plotone{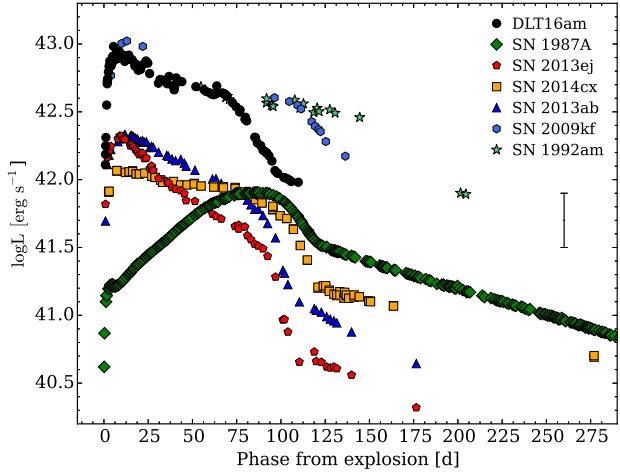}
\caption{Pseudo-bolometric light curve of DLT16am compared to those of other Type II SNe. Luminosities were obtained using similar filters and integration limits. A representative error bar is also shown, corresponding to the uncertainty on the derived extinction in the direction of DLT16am. \label{fig:bolometric}}
\end{center}
\end{figure}

The pseudo-bolometric light curve of DLT16am was computed converting the available magnitudes to flux densities at the corresponding effective wavelengths, subsequently integrating using Simpson's rule.
The resulting light curve is shown in Figure~\ref{fig:bolometric}, along with those of other SNe computed following the same prescriptions and using similar filters. 
From the analysis of the bolometric light curve of DLT16am, we infer a peak luminosity of $\log{L}\simeq43\,\rm{erg^{-1}}\,\rm{s}$, which has to be considered a lower limit, since the contribution of the UV flux is, in general, important in early SN light curves. 
On the other hand, at later times (e.g. on the radioactive-decay tail) UV bands give a minor contribution to the total flux.
We can therefore use the pseudo-bolometric light curve to infer the $^{56}\rm{Ni}$ mass produced during the SN explosion.
This quantity is generally estimated using the method described in \citet{2014MNRAS.439.2873S}, taking the bolometric luminosity of SN~1987A during the nebular phase as a reference.

We therefore extrapolated the pseudo-bolometric light curve of DLT16am assuming a complete trapping of the $\gamma$-rays produced by the $^{56}$Co decay ($\simeq1\,\rm{mag}/100\,\rm{d}$).
Starting from the last observed point (which is likely a few days after the onset of the radioactive tail), we compared the extrapolated luminosity at $\simeq+150\,\rm{d}$ with that of SN~1987A computed using the same integration limits and at the same phase, to get a rough estimate of the mass of radioactive $^{56}\rm{Ni}$ deposited in the SN ejecta.
Using the relation:
\begin{equation}
M(^{56}\mathrm{Ni})=0.075\mathrm{M_{\odot}}\times L_{SN}(t)/L_{87A}(t)),
\end{equation}
we infer a relatively high $^{56}\rm{Ni}$ mass of $0.208\pm0.044$\msun.
A similar amount of radioactive $^{56}\rm{Ni}$ was derived for the Type II-P SN~1992am \citep{1994AJ....107.1444S}, while a larger limit is given for SN~2009kf \citep[$M_{^{56}Ni}<0.4$\msun; see][]{2010ApJ...717L..52B}.
Although the derived luminosity might be significantly affected by the large uncertainty on the estimated extinction, we remark that the photospheric expansion velocity inferred from the spectroscopic analysis is in agreement with the derived bright absolute $i$-band magnitude (and hence the derived total reddening) for DLT16am, according to the existing luminosity-velocity relation \citep{2002ApJ...566L..63H} for SNe II (see Section~\ref{sec:comparison}).
In addition, also the derived $^{56}$Ni mass seems to be in agreement with the trend followed by Type II SNe with comparable photospheric velocities and absolute magnitudes \citep[see][]{2003ApJ...582..905H}.

\section{Spectroscopy} \label{sec:spectroscopy}
Our spectroscopic follow-up campaign started on 2016 November 22.64~UT ($\simeq2\,\rm{d}$ after the explosion), lasting up to 2017 March, 7.01~UT (at $\simeq110\,\rm{d}$).
NIR spectra were obtained within the same period, with a lower cadence.
The flux calibration was checked using photometric information obtained during the closest nights, scaled using low order polynomials.

\subsection{Optical spectra} \label{sec:opt_spectra}
Early spectra are dominated by a nearly featureless and very red continuum, with narrow \ha~and \ion{[S}{2]} lines from the host and little or no flux at wavelengths shorter than 5000\ang. 
After correcting for the foreground Galactic extinction ($A_V=0.042\,\rm{mag}$), redshift ($z=0.00456$, inferred from the X-shooter spectrum; see Section~\ref{sec:host}) and host extinction ($A_V\simeq6.12\,\rm{mag}$), we estimated the temperature of the ejecta at early phases using simple black-body fits to the spectral continuum through the \textsc{curve{\_}fit} tool available in the \textsc{SciPy}\footnote{\url{https://www.scipy.org/}} package \citep{scipy,2011arXiv1102.1523V}. 
We find a rapid decrease, from $16300\pm6600\,\rm{K}$ at $+2\,\rm{d}$ to $10600\pm3200\,\rm{K}$ at $+5\,\rm{d}$, in agreement with the fast color evolution within the early phases (see Section~\ref{sec:photometry}), although the large uncertainties might suggest a lack of clear evolution in temperature.
The errors on the derived temperatures were estimated by applying different extinction values, within the range of the derived uncertainty on the reddening (see Section~\ref{sec:reddening}).
At later phases ($t>+5\,\rm{d}$), we could not determine the temperatures, since no contribution from the SN to the spectral continuum at $\lambda<6000$\ang~was observed due to the high extinction (see Section~\ref{sec:reddening}).
\begin{figure*}
\begin{center}
\epsscale{1.15}
\plotone{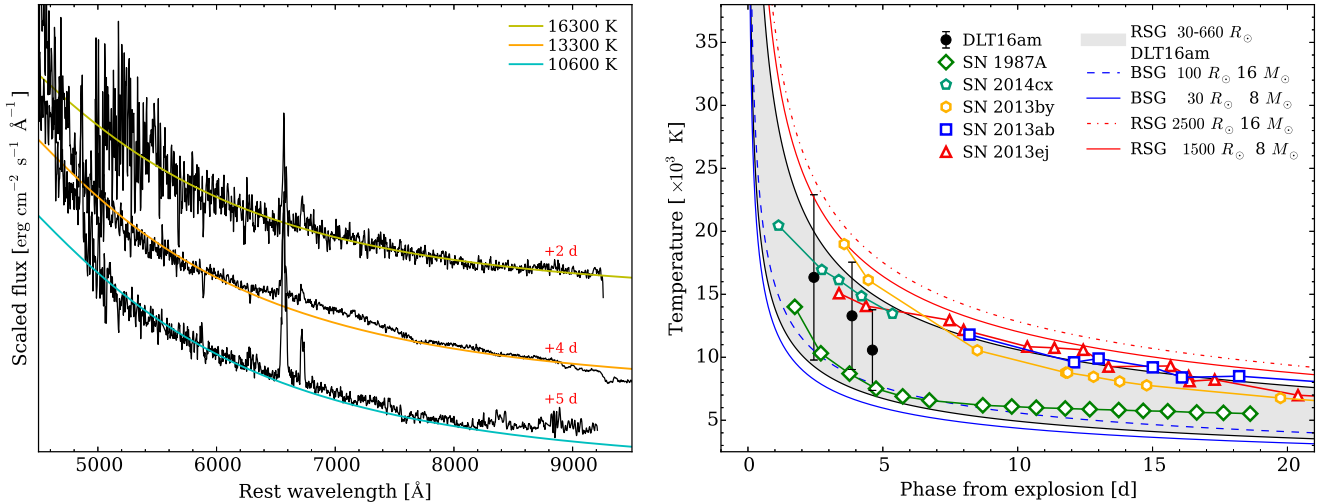}
\caption{{\bf Left:} Black-body fits to the early spectra of DLT16am. The spectra have been corrected for the derived total extinction in the direction of DLT16am. {\bf Right:} Progenitor radius of DLT16am contrained using the formalism of \citet{2011ApJ...728...63R}. Models for RSGs and BSGs (red and blue lines, respectively) are also shown, while the range of radii obtained for DLT16am is delimited by black lines. Temperature evolutions obtained for other Type II SNe are also shown for comparison. \label{fig:tempradius}}
\end{center}
\end{figure*}

Figure~\ref{fig:tempradius} (left panel) shows the results of the black body fit to the spectra up to $\simeq+5\,\rm{d}$.
As discussed by \citet{2011ApJ...728...63R}, the time scale of the cooling phase subsequent to the SN shock breakout heating of the progenitor envelope, strongly depends on the initial progenitor properties, such as its radius, density profile, opacity and composition.
Using their formalism (i.e. their equation 12), we infer a rough estimate of the progenitor radius fitting the temperature evolution of DLT16am during the first $\simeq5\,\rm{d}$ of its spectroscopic evolution \citep[see][]{2011ApJ...728...63R} and \citep[][for an analysis on the limitations of analytic models]{2017ApJ...848....8R}.
Figure~\ref{fig:tempradius} (right panel) shows the resulting fit (obtained assuming an explosion energy of $10^{51}\,\rm{erg}$ and a typical optical opacity for a H-rich gas, $\kappa=0.34\,\rm{cm^2}\,\rm{g^{-1}}$; \citealt{2011ApJ...728...63R}), along with the temperature evolution obtained for other Type II SNe.

Direct imaging in deep pre-SN archival images confirmed the claim that the majority of SNe II have red super-giant (RSG) progenitors \citep[see, e.g.,][]{2005MNRAS.364L..33M,smartt09,2011MNRAS.417.1417F,2012ApJ...756..131V,2012AJ....143...19V}. 
We therefore adopt the typical mass range for RSG progenitors \citep[$8-17$\msun;][]{smartt09}, and using the \citet{2011ApJ...728...63R} formalism we obtain a radius of $30-660$\rsun~for the progenitor of DLT16am, where the uncertainty is largely due to the error on the estimated reddening. 
Although the the model is weakly dependent on the mass of the progenitor \citep[see][]{2011ApJ...728...63R}, the large uncertainty on the derived temperatures does not allow us to rule out a blue super-giant (BSG) star as a viable progenitor for DLT16am.

From $+24\,\rm{d}$ a broad Full-Width-at-Half-Maximum - $\rm{FWHM}\simeq9000$\kms) \ha~feature in emission starts to dominate the flux, masking the presence of the host \ion{[S}{2]} lines, with blue-shifted peaks typical of Type II SNe \citep[see][for a discussion]{2014MNRAS.441..671A}.
Following \citet{2014MNRAS.441..671A} and \citet{2014ApJ...786L..15G}, we measure the \ha~blue-shifted emission offset and the ratio between the EWs of its absorption and emission P-Cygni components ($a/e$).
We find a significant blue-shifted \ha~peak ($V\simeq3300$\kms) at $+30\,\rm{d}$ and a small contribution of the absorption component to the \ha~P-Cygni profile ($a/e\simeq0.06$), both indicative of fast declining light curves during the plateau phase and in agreement with the results of our photometric analysis (see Section~\ref{sec:comparison}).
From the same phase we also detect the NIR (8498, 8542, 8662\ang) \ion{Ca}{2} and the \ion{O}{1} (7772, 7774, 7775\ang) triplets, both becoming more evident at later phases.
\ion{C}{1} $\lambda9095$ and \ion{Mg}{2} $\lambda9218$ lines appear between $+58$ and $+59\,\rm{d}$, while forbidden \ion{[Ca}{2]} lines are visible from $+91\,\rm{d}$, marking the onset of the nebular phase, possibly blended with the \ion{[O}{2]} 7319, 7330\ang~doublet. 
From the same phase we note a significant change in the relative strengths of the \ion{Ca}{2} NIR triplet, which is likely caused by the appearance of the nebular \ion{[O}{1]} 8446\ang~line.

\begin{figure*}
\begin{center}
\epsscale{1.18}
\plotone{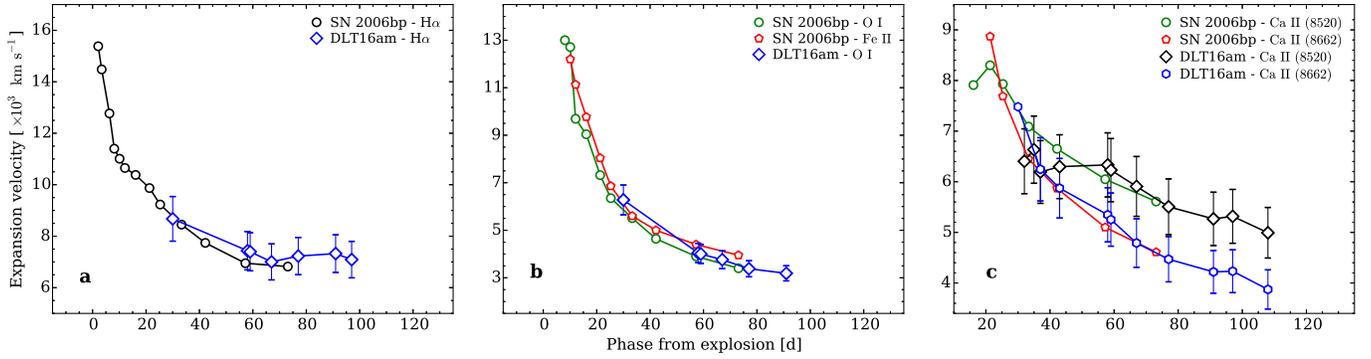}
\caption{Expansion velocity evolution derived from \ha~{\bf (a)}, \ion{O}{1} $\lambda7773$ and \ion{Fe}{2} $\lambda5169$ {\bf (b)} and the NIR \ion{Ca}{2} triplet {\bf (c)}, compared to those derived for SN~2006bp. The comparison is made on the basis of the best match of the DLT16am $+59\,\rm{d}$ spectrum obtained with the \textsc{SNID} tool. \label{fig:vel_comparison}}
\end{center}
\end{figure*}

From the positions of the minima of the P-Cygni absorption profiles we derived an estimate of the expansion velocities for different ions.
\ha~P-Cygni profiles are clearly visible only from $+30\,\rm{d}$, when we derive an expansion velocity of $\simeq8700$\kms, slowly declining to $\simeq7100$\kms~at $+97\,\rm{d}$ (see also Figure~\ref{fig:vel_comparison}, panel a).
While the \ion{Fe}{2} $\lambda5169$ line is generally assumed as a good tracer of the photospheric velocity, we could not detect this line in any of our spectra of DLT16am due to the high extinction (see Section~\ref{sec:reddening}).
We therefore obtained a rough estimate of the photospheric velocity from the minima of the prominent \ion{O}{1} triplet, getting a relatively fast evolution from $\simeq6300$\kms~(at $+30\,\rm{d}$) to $\simeq3000$\kms~(at $+97\,\rm{d}$; see Section~\ref{sec:comparison} and Figure~\ref{fig:vel_comparison} for a detailed discussion).
The \ion{Ca}{2} NIR triplet was partially resolved, and we inferred an expansion velocity declining from $\simeq6400$\kms~to $5000\simeq$\kms~for the blend of the 8498,8542\ang~lines, and from $7500$\kms~to $3900$\kms~for the 8662\ang~line.

\subsection{Near infra-red spectra} \label{sec:nir_spectra}
At $+7\,\rm{d}$, our SOFI spectrum shows an almost featureless continuum, while at later phases spectra show a significant metamorphosis, as broad H lines with prominent P-Cygni profiles start to dominate the flux.
Following the NIR line identifications proposed for SNe~2004et \citep{2010MNRAS.404..981M} and 1998S \citep{2004MNRAS.352..457P}, we identify most of the main H Paschen (from $\rm{Pa}\alpha$ up to $\rm{Pa}\zeta$) and Brackett (from $\rm{Br}\gamma$ up to $\rm{Br}\eta$) lines, although the $\rm{Pa}\gamma$ line is most likely a blend with \ion{He}{1} (10830\ang).
Beginning on $+34\,\rm{d}$, we also detect the \ion{Ca}{2} NIR triplet. We also identify \ion{Sr}{2} $\lambda10327$ in the blue part of the spectra beginning on $+43\,\rm{d}$, when the feature is still partially blended with the Pa$\delta$ line, becoming more evident at later phases.

\ion{C}{1} and \ion{Mg}{2} lines appear from $+54\,\rm{d}$, in agreement with the analysis performed on the optical spectra (although we cannot rule out the presence of these lines also at $+43\,\rm{d}$).
From the $\rm{Pa}\beta$ absorption minima, we infer an expansion velocity decreasing from $\simeq8340$\kms~at $+34\,\rm{d}$ to $\simeq3460$\kms~at $+107\,\rm{d}$, which is consistent with the evolution derived from \ha~(see above), while measuring the prominent blue minimum in the blue part of the $\rm{Pa}\gamma$+\ion{He}{1} 10830\ang~profile, we infer higher expansion velocities, declining from $\simeq9000$\kms~to $\simeq5660$\kms~from $+34\,\rm{d}$ to $+107\,\rm{d}$, respectively.

\begin{figure}
\begin{center}
\epsscale{1.15}
\plotone{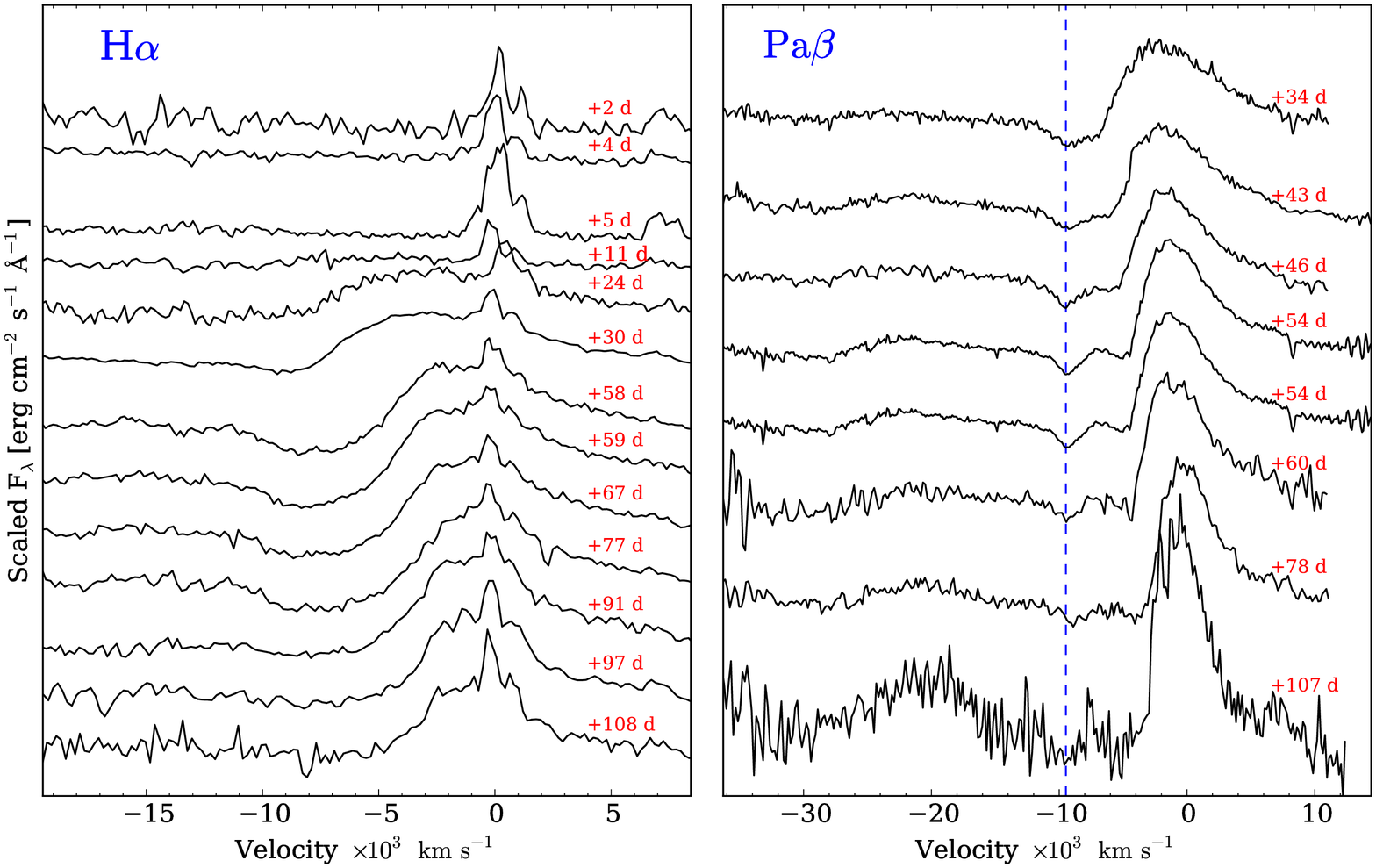}
\caption{Details of the evolution of the \ha~and Pa$\beta$ profiles of DLT16am in the velocity space. The position of the blue absorption feature in the $\rm{Pa}\beta$ profile, with a constant expansion velocity of $9400-9500$\kms~is marked with a dashed blue line. Phases refer to the epoch of the explosion. Spectra with lower SNR are not included. \label{fig:highvel_components}}
\end{center}
\end{figure}

From $+43\,\rm{d}$ we notice a second absorption feature in the blue part of the P-Cygni absorption component of $\rm{Pa}\beta$, becoming more evident at later phases, which might be attributed to a rapidly expanding outer H shell moving at a roughly constant velocity ($9400-9500$\kms).
High velocity (HV) components not evolving in time were observed also in the \ha~profile and \ion{He}{1} 10830\ang~of other SNe II (see, e.g., the case of SNe~2009bw; \citealt{2012MNRAS.422.1122I}, 1999em and 2004dj; \citealt{2012ApJ...761..100C} and the discussion in \citealt{2007ApJ...662.1136C}) and attributed to weak interaction between the SN ejecta and a pre-existing circumstellar medium surrounding the progenitor star. 
Analyzing HV components in the \ha~and \ion{He}{1} 10830\ang~profiles during the photospheric phase, \citet{2007ApJ...662.1136C} computed a model for ejecta - circumstellar interaction in SNe II-P.
Although the expansion velocities inferred from the Pa$\gamma$ P-Cygni minimum suggest that the entire absorption component might be mainly attributed to the \ion{He}{1} 10830\ang~line, Figure~\ref{fig:highvel_components} shows that a similar feature was never observed in the \ha~profile at any phase, with the possible exception of the $+58\rm{d}$ spectrum.
On the other hand, the signal-to-noise ratio (SNR) of our spectra is not sufficient to safely rule out the presence of this feature at optical wavelengths.
\begin{figure*}
\begin{center}
\epsscale{1.15}
\includegraphics[width=\columnwidth]{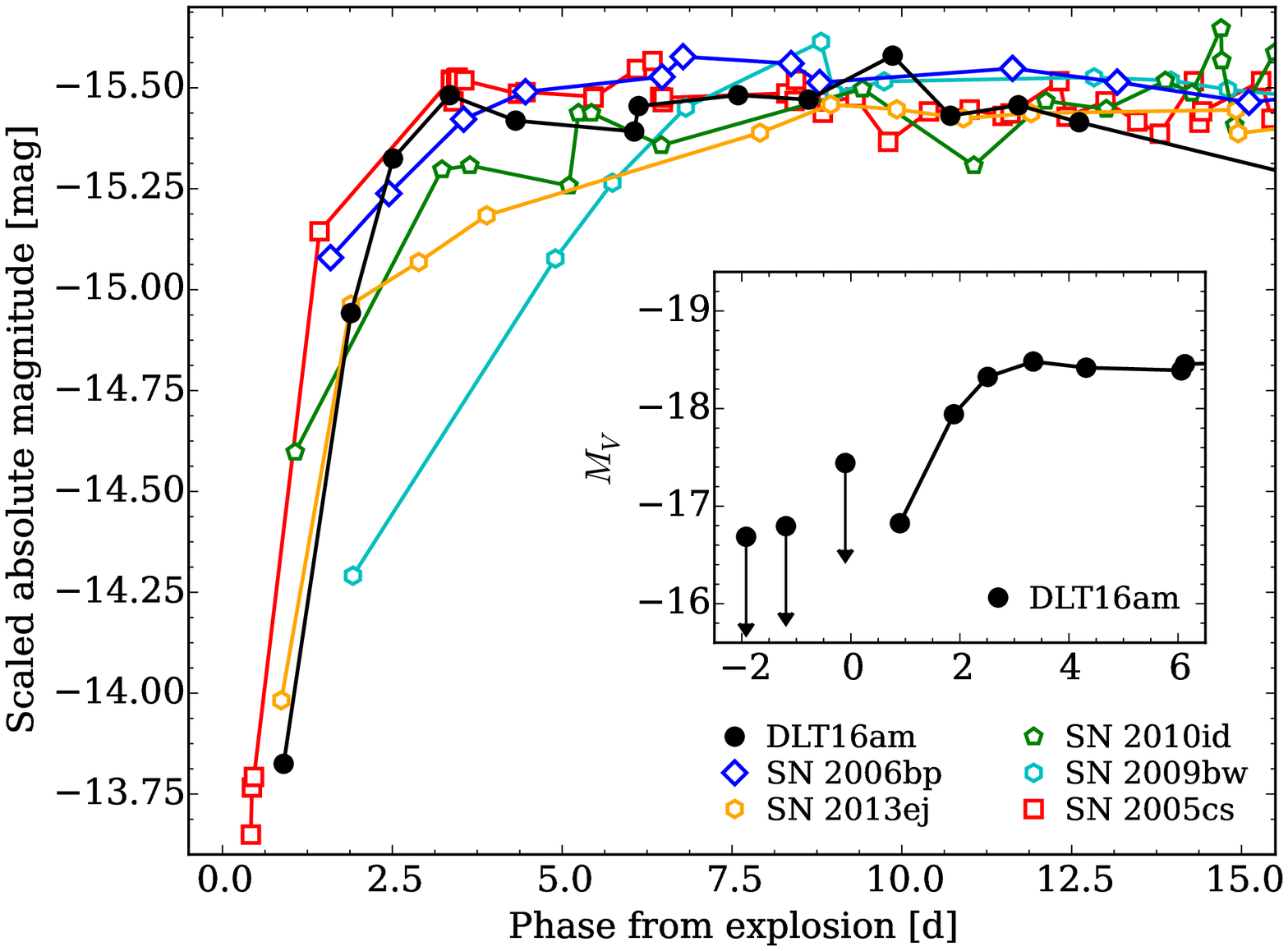}
\includegraphics[width=0.515\linewidth]{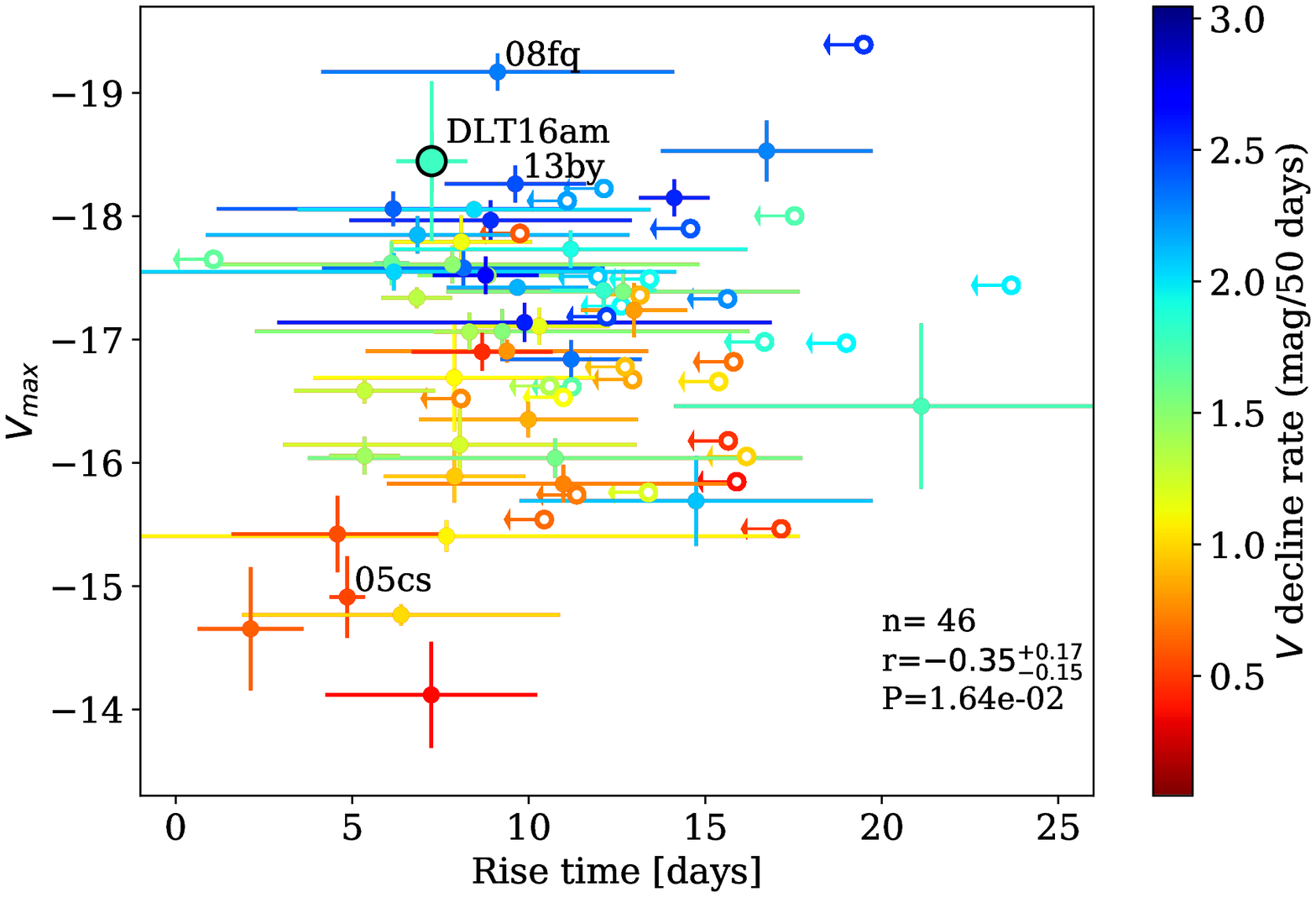}
\caption{{\bf Left:} Early absolute $V$-band light curve of DLT16am compared to other Type II SNe (the $r$-band light curve of SN~2010id is shown for comparison). The light curves are scaled to the plateau luminosity of SN~2010id. Phases refer to the epochs of the explosions. {\bf Right:} $V$-band absolute magnitude versus rise time for the same sample of SNe II, color-coded according to their $V$-band decline rate ($S_{50V}$). $n$ is the number of transients in the plot, $r$ the Pearson $s$-correlation coefficient and $P$ the probability of detecting a correlation by chance, as in \citet{valenti16}. \label{fig:vband_results}}
\end{center}
\end{figure*}

\section{Comparison with other Type II SNe} \label{sec:comparison}
In Figure~\ref{fig:abs_colors} we have compared the main photometric properties of a sample of luminous and more canonical Type II SNe.
Left panels show that DLT16am is brighter than the normal Type II SNe~2014cx \citep{2016ApJ...832..139H} and 2013ej \citep[][although the large uncertainties on the extinction suggest comparable plateau magnitudes]{valenti14}, while high luminosities are not unusual among SNe II \citep[see, e.g., the case of SN~2013by;][]{2015MNRAS.448.2608V}.

Following the prescriptions of \citet{valenti16}, we estimate photometric parameters using the $V$-band photometric evolution.
As $V$-band maximum we consider the point at which the variation in magnitude is less than $0.1\,\rm{mag}\,\rm{d^{-1}}$, while $S_{50V}$, the decline rate in $\rm{mag}/50\,\rm{d}$ was computed soon after the maximum ($10\,\rm{d}$ after the explosion) to $+50\,\rm{d}$ after the explosion.
Figure~\ref{fig:vband_results} summarize the results of this analysis, compared to those obtained for other SNe II \citep{valenti16}.
The $V$-band maximum for DLT16am occurs on $\rm{JD}=2457719.95$, corresponding to an absolute magnitude $M_V=-18.49\pm0.65\,\rm{mag}$ (where the error is almost entirely due to the uncertainty on the reddening) and to a relatively fast rise time of $7.4\pm1.0\,\rm{d}$.
At redder wavelengths we find longer rises (10.9, 11.4 and $11.9\,\rm{d}$ in $r$-, $i$- and $z$-band, respectively), in agreement with the results obtained by \citet{2015MNRAS.451.2212G} on a large sample of SNe II.
In {\it r} band, on the other hand, we find that DLT16am has a relatively slow rise if compared to the sample of \citet{2016ApJ...820...33R}

Comparing the early-time absolute $r$-band light curve of SN~2010id with those of SNe~2005cs \citep{2009MNRAS.394.2266P} and SN~2006bp \citep{2007ApJ...666.1093Q}, \citet{2011ApJ...736..159G} suggested a possible trend for sub-luminous and `normal' events, with faster rises with a sharp onset of the flat plateau for fainter objects, with SN~2006bp showing a more gradual transition over a longer period.
Including the slow-rising SNe~2009bw \citep{2012MNRAS.422.1122I} and 2013ej $R$-band light curves in the comparison, \citet{valenti14} also argued that slow rising Type II SNe might be brighter than fast rising transients.
The fast rise and the bright absolute magnitude shown by DLT16am seem to contradict these predictions, confirming the results obtained by \citet{2016ApJ...820...33R} on a sample of $R$-band light curves of Type II SNe.
In Figure~\ref{fig:vband_results}, left panel, we compare the early $V$-band absolute light curve of DLT16am with those of a sample of subluminous (SNe~2005cs; \citealt{2009MNRAS.394.2266P} and 2010id; \citealt{2011ApJ...736..159G}), normal (SNe~2006bp; \citealt{2007ApJ...666.1093Q} and 2013ej \citealt{valenti14}) and luminous (SN~2013by; \citealt{2015MNRAS.448.2608V} and DLT16am) SNe II, all scaled to the luminosity of the plateau of SN~2010id, in order to gain a better insight into their different rise times. 
We find a particularly good match with the sub-luminous SN~2005cs, while more luminous transients, like SNe~2006bp and 2013ej, seem to have significantly longer rises to maximum light.
This trend is confirmed by the comparison with a larger sample of SNe II shown in Figure~\ref{fig:vband_results} (right panel).
DLT16am lies close to the brighter end of the absolute peak magnitude range, with a rise time comparable to those displayed by the faintest objects.
Comparing the $r$-band absolute peak magnitude and rise time to the sample of \citet{2016ApJ...820...33R} (see their Figure 10), on the other hand, we find a longer rise time, while DLT16am falls in a scarcely populated region of their luminosity-rise time diagram.
On the other hand, we have to remark that this might be due to the lack of transients with well constrained explosion epochs (see, e.g., the case of SN~2008fq and the large uncertanties in the rise times of the brightest objects).

\begin{figure}
\begin{center}
\epsscale{1.15}
\plotone{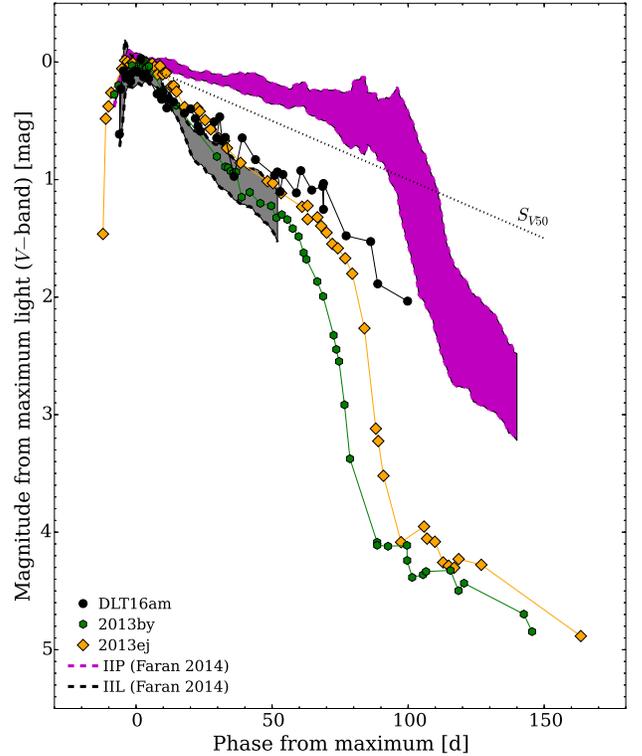}
\caption{Absolute $V$-band light curve of DLT16am and other similar transients compared with templates computed by \citet{2014MNRAS.442..844F,2014MNRAS.445..554F} for Type II SNe.
The dotted black line marks the limiting slope ($S_{50V}$; see the text for details) between Type II-P (magenta dashed lines) and II-L (black dashed models) SNe. \label{fig:16am_templ}}
\end{center}
\end{figure}

The historical classification of Type II in II-P and II-L SNe has recently been a matter of debate \citep[see, e.g.,][]{2012ApJ...756L..30A,2014ApJ...786...67A,2014MNRAS.442..844F,2014MNRAS.445..554F,2015ApJ...799..208S,valenti16}.
In order to give an accurate classification of DLT16am, in Figure~\ref{fig:16am_templ} we compare the $V$-band light curve of DLT16am to the Type II-P and II-L SNe templates computed by \citet{2014MNRAS.442..844F,2014MNRAS.445..554F}.
Although DLT16am shows photometric features typically observed in Type II-P SNe (namely an extended plateau after maximum, with a subsequent steep drop in magnitude around $+80\,\rm{d}$), like SN~2013by \citep{valenti14} its $V$-band light curve lies close to the bright end of the II-L templates, in an intermediate region between II-L and II-P templates.
This is in agreement with the decline rate derived from the $V$-band light curve ($S_{50V}=0.84\pm0.04\,\rm{mag}/50\,\rm{d}$), which, according to \citet{2014MNRAS.445..554F}, is greater than the limit for Type II-P SNe ($0.5\,\rm{mag}/50\,\rm{d}$; see Figure~\ref{fig:16am_templ}).
Following \citet{2016ApJ...828..111R} and their proposed classification based on the early light curves morphology, we compare the $r$-band light curve of DLT16am to the results of their analysis on the sample of \citet{2016ApJ...820...33R}, obtaining a good match with their fast rise--fast decline (II-FF) cluster of Type II SNe.

\begin{figure}
\begin{center}
\epsscale{1.15}
\plotone{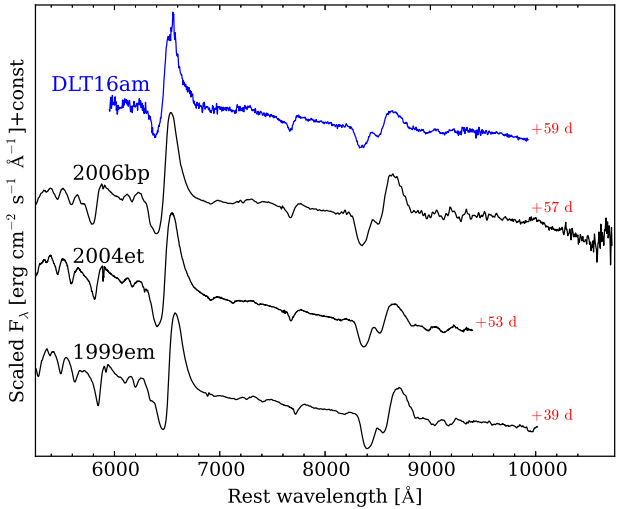}
\caption{Comparison of the $+59\,\rm{d}$ spectrum of DLT16am with those of a sample of Type II SNe at similar phases. Comparison objects were selected on the basis of the results obtained using the SNID comparison tool \citep{2007ApJ...666.1024B}. Fluxes have been scaled to arbitrary constants. \label{fig:spec_comparison}}
\end{center}
\end{figure}

Figure~\ref{fig:spec_comparison} shows a comparison of our $+59\,\rm{d}$ spectrum with those of other Type II SNe, based on the best fits of the spectral features to archival spectra obtained using the `Supernova Identification' \citep[SNID\footnote{\url{https://people.lam.fr/blondin.stephane/software/snid/}};][]{2007ApJ...666.1024B}) tool.
While the best match was obtained with the Type II-P SN~2006bp \citep{2007ApJ...666.1093Q,2008ApJ...675..644D}, good fits of the spectral features were obtained also with the Type II-P SNe~2004et \citep{2006MNRAS.372.1315S,2007MNRAS.381..280M,2010MNRAS.404..981M} and 1999em \citep{2001ApJ...558..615H,2002PASP..114...35L,2003MNRAS.338..939E,2006A&A...447..691D}.
Based on this similarity, we compared the expansion velocities of DLT16am with those obtained for SN~2006bp, obtaining similar values for all the ions visible in both set of spectra (see Figure~\ref{fig:vel_comparison}).
Due to the high extinction, we could not compare the expansion velocities inferred from \ion{Fe}{2} (5169\ang) or \ion{Sc}{2} (6246\ang), which are typically considered good indicators of the photospheric velocity.
On the other hand, SN~2006bp shows similar \ion{Fe}{2} (5169\ang) and \ion{O}{1} (7773\ang) expansion velocities (see Figure~\ref{fig:vel_comparison}, panel b) and based on the strong spectroscopic similarities between the two transients, we can therefore use the velocity evolution inferred from the \ion{O}{1} minima as a rough estimate of the photospheric expansion velocity for DLT16am.
Using the existing luminosity-velocity relation for Type II SNe \citep{2002ApJ...566L..63H,2003astro.ph..9122H} we can therefore perform an independent consistency check on the derived host galaxy reddening in the direction of DLT16am (see Section~\ref{sec:reddening}).
\citet{2002ApJ...566L..63H} use $+50\,\rm{d}$ as an indicative epoch (roughly the mid point of the plateau phase) and, although DLT16am shows a relatively short plateau lasting $\simeq80\,\rm{d}$, we will adopt the same approach, comparing the expansion velocity and absolute $V$-magnitude with those obtained from the sample of \citet{2003astro.ph..9122H} at similar phases. 
A similar approach was adopted also for the absolute $V$-band luminosity, where we took the uncertainty on the distance modulus and the total extinction as an estimate of the error on the derived magnitude.
In Figure~\ref{fig:lumVelRel} we compare the results for DLT16am with those obtained for the sample of \citet{2003astro.ph..9122H}.
With a photospheric velocity of $\simeq4585$\kms~and  an absolute $V$-band magnitude at $+50\,\rm{d}$ of $M_V=-17.40\pm0.65\,\rm{mag}$, DLT16am falls in the region of other luminous SNe II, in agreement with the expectation that luminous transients have higher expansion velocities \citep[see also][]{2016ApJ...820...33R}.
The correlation between absolute peak magnitudes and expansion velocity was recently confirmed by \citet{2017arXiv170902799G}, who also confirmed the previous claim of \citet{2014MNRAS.441..671A} and \citet{valenti16} that brighter SNe II show shorter plateau phases and steeper decline rates.
Similar results were also reported by \citet{2016AJ....151...33G} analyzing the light curves of a large sample of Type II SNe.

\section{Summary and Conclusions}  \label{sec:conclusions}
We have discussed the results of our analysis on the photometric and spectroscopic data obtained during our follow-up campaign of the Type II SN DLT16am (aka SN~2016ija).
The transient was discovered during the ongoing DLT40 survey, which is monitoring a sample of galaxies in the nearby Universe in search for young SNe within the first days from the explosion.

Early spectra showed a highly reddened, nearly featureless continuum, while H, \ion{Ca}{2} and \ion{O}{1} lines with prominent P-Cygni profiles gradually appear at later phases.
The comparison of the colors of DLT16am with those of other similar transients during the plateau phase suggests a contribution of $E(B-V)=1.95\pm0.15\,\rm{mag}$ from the host galaxy (NGC~1532) to the total extinction (see Section~\ref{sec:reddening} and Figure~\ref{fig:abs_colors}).
Although DLT16am was clearly detected by our survey at redder optical wavelengths, its substantial extinction of $Av\simeq6\,\rm{mag}$ supports the claim that optical surveys might be missing a significant fraction of nearby, highly reddened SNe \citep[e.g.][]{2012ApJ...756..111M,2017ApJ...837..167J}. 
Other examples of recent heavily obscured CCSNe observed in nearby galaxies include SN~2009hd \citep{2011ApJ...742....6E}, SN~2005at \citep{2014A&A...572A..75K} and SN 2013fc \citep{2016MNRAS.456..323K}, SPIRITS~15c and SPIRITS~14buu; \citealt{2017ApJ...837..167J}, or SNe~2008cs, 2011hi, and 2010P \citealt{2008ApJ...689L..97K,2012ApJ...744L..19K,2014MNRAS.440.1052K}). 
Such events can have important implications for the comparison between CCSN rates and the cosmic star formation history.

Assuming a standard ($R_V=3.1$) extinction law \citep{1989ApJ...345..245C} and a distance modulus $\mu=31.51\pm0.20\,\rm{mag}$ \citep{2013AJ....146...86T}, we obtain a relatively bright absolute peak magnitude ($M_V=-18.49\pm0.65\,\rm{mag}$) compared to those displayed by other Type II-P-like SNe.

\begin{figure}
\begin{center}
\epsscale{1.15}
\plotone{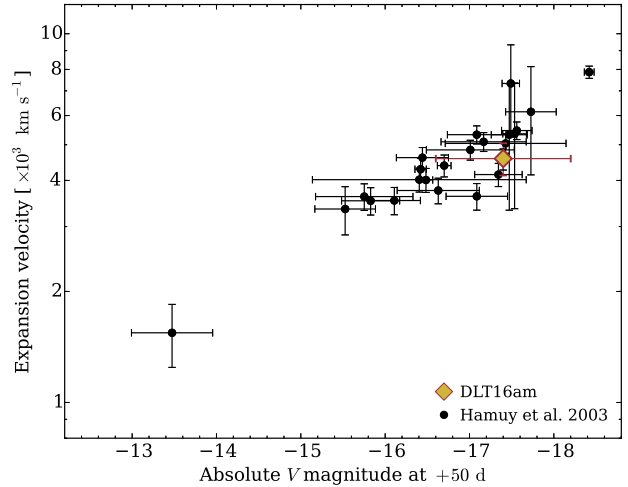}
\caption{Expansion velocity (inferred from \ion{Fe}{2} $\lambda5169$ absorption minima) versus $V$-band absolute magnitude for a sample of SNe II. Velocities and luminosities have been computed $50\,\rm{d}$ after the explosion \citep[roughly the mid-point of the plateau phase; see][]{2001ApJ...558..615H}. DLT16am is marked with a yellow and red diamond. \label{fig:lumVelRel}}
\end{center}
\end{figure}

The absolute magnitude at $+50\,\rm{d}$ is consistent with the photospheric velocity inferred at the same phase, according to the existing luminosity-velocity relation for Type II SNe \citep[see][and Figure~\ref{fig:lumVelRel}]{2002ApJ...566L..63H,2003astro.ph..9122H}. 
The derived slope within $50\,\rm{d}$ from the explosion ($S_{50V}$; see Section~\ref{sec:comparison}) suggests a relatively steep decline during the plateau phase ($S_{50V}=0.84\pm0.04\,\rm{mag}/50\,\rm{d}$), which, according to \citet{2014MNRAS.445..554F} (who give a limit of $0.5\,\rm{mag}/50\,\rm{d}$ to the maximum slope for SNe II-P), means that DLT16am should be considered a Type II-L SN.
A similar conclusion was reached by \citet{valenti14} in their analysis of SN~2013by, which, like DLT16am, showed an extended plateau, with a drop in magnitude around $80\,\rm{d}$ after the explosion (see also Figure~\ref{fig:abs_colors}).

In Figure~\ref{fig:vband_results} we have shown a comparison of the fundamental photometric parameters inferred for DLT16am with those obtained for a sample of SNe II.
The fast rise time ($7.4\pm1.0\,\rm{d}$) and the bright $V$-band absolute magnitude at maximum ($M_V=-18.49\pm0.65\,\rm{mag}$) seem to contradict the prediction that luminous Type II SNe have long rise times compared to those of sub-luminous events.
The lack of other Type II SNe in the region where DLT16am falls highlights its peculiarity, but might also be due to the lack of transients with well constrained explosion epochs.

It is therefore crucial to increase our sample of SNe II with well constrained explosion epochs and rise times, providing early discoveries and subsequent multi-wavelength data.
With its estimated rate of $\simeq4-5$ SNe II per year discovered within $1\,\rm{d}$ from the explosion, confirmed by the number of discoveries during the first year of operations, the DLT40 survey will significantly increase the number of early discoveries, helping to further explore correlations between fundamental parameters of Type II SNe.

\acknowledgments
\noindent Based on observations collected at:
The Very Large Telescope operated by the European Organisation for Astronomical Research in the Southern hemisphere, Chile as part of the ESO large programme 198.A-0915.. \\
ESO La Silla Observatory as part of PESSTO (197.D-1075.191.D-0935). \\ 
The Gemini Observatory, under program GN-2016B-Q-57, which is operated by the Association of Universities for Research in Astronomy, Inc., under a cooperative agreement with the NSF on behalf of the Gemini partnership: the National Science Foundation (United States), the National Research Council (Canada), CONICYT (Chile), Ministerio de Ciencia, Tecnolog\'{i}a e Innovaci\'{o}n Productiva (Argentina), and Minist\'{e}rio da Ci\^{e}ncia, Tecnologia e Inova\c{c}\~{a}o (Brazil). \\
Based on observations made with the Nordic Optical Telescope, operated by the Nordic Optical Telescope Scientific Association at the Observatorio del Roque de los Muchachos, La Palma, Spain, of the Instituto de Astrofisica de Canarias. \\
This paper includes data gathered with the 6.5 meter Magellan Telescopes located at Las Campanas Observatory, Chile. \\
This work makes use of observations from the Las Cumbres Observatory network of telescopes. \\
This work includes data obtained with the Swope Telescope at Las Campanas Observatory, Chile, as part of the Swope Time Domain Key Project (PI Piro, Co-PIs Shappee, Drout, Madore, Phillips, Foley, and Hsiao).

SNOoPy is a package for SN photometry using PSF fitting and/or template subtraction developed by E.~Cappellaro. A package description can be found at \url{http://sngroup.oapd.inaf.it/snoopy.html}. \\
\textsc{Foscgui} is a graphic user interface aimed at extracting SN spectroscopy and photometry obtained with FOSC-like instruments. It was developed by E.~Cappellaro. A package description can be found at \url{http://sngroup.oapd.inaf.it/foscgui.html}.

Research by D.J.S. and L.T. l is supported by NSF grant AST-1412504 and AST-1517649. \\
C.G acknowledges support from the Carlsberg Foundation. \\
R.C. and M.S. acknowledge support from STFC grant ST/L000679/1 and EU/FP7-ERC grant no [615929]
T.W.C. acknowledges the support through the Sofia Kovalevskaja Award to P. Schady from the Alexander von Humboldt Foundation of Germany. \\
Part of the funding for GROND (both hardware as well as personnel) was generously granted from the Leibniz-Prize to Prof. G. Hasinger (DFG grant HA 1850/28-1). \\
Support for G.P. is provided by the Ministry of Economy, Development, and Tourism's Millennium Science Initiative through grant IC120009, awarded to The Millennium Institute of Astrophysics, MAS. \\
J.H. acknowledges financial support from the Finnish Cultural Foundation and the Vilho, Yrj\"o and Kalle V\"ais\"al\"a Foundation of the Finnish Academy of Science and Letters. \\
D.A.H., C.M., and G.H. are supported by NSF grant 1313484. \\
M.D.S. acknowledges support by a research grant (13261) from the VILLUM FONDEN and for financial support of NUTS by the the Instrument Center for Danish Astrophysics (IDA). \\
L.G. was supported in part by the US National Science Foundation under Grant AST-1311862. \\
M.M.P., N.M. and E.Y.H. acknowledge the support provided by the National Science Foundation under Grant No. AST-1008343, AST-1613472 and AST-1613426. \\
A.G.-Y. is supported by the EU via ERC grant No. 725161, the Quantum Universe I-Core program, the ISF, the BSF Transformative program and by a Kimmel award. \\
K.M. acknowledges support from the STFC through an Ernest Rutherford Fellowship. \\
The UCSC group is supported in part by NSF grant AST--1518052, the Gordon \& Betty Moore Foundation, and from fellowships from the Alfred P.\ Sloan Foundation and the David and Lucile Packard Foundation to R.J.F.

\facilities{VLT:Kueyen \citep[X-shooter spectrograph][]{2011A&A...536A.105V}, NTT (EFOSC2 \citealt{1984Msngr..38....9B} and SOFI \citealt{1998Msngr..91....9M} spectrographs), Gemini:Gillett \citep[GNIRS spectrograph;][]{GNIRS}, NOT (ALFOSC and NOTCam cameras), Magellan:Baade (FIRE spectrograph), FTN, FTS (FLOYDS spectrographs), LCOGT, Swope, Liverpool:2m (IO:O camera), Max Planck:2.2m (GROND camera), CTIO:PROMPT	 \citep[PROPT5 telescope;][]{prompt}}

\software{SNID \citep{2007ApJ...666.1024B}, Foscgui (\url{http://graspa.oapd.inaf.it/foscgui.html}), SNOoPy (\url{http://sngroup.oapd.inaf.it/snoopy.html}), PySALT \citep{2010SPIE.7737E..25C}, XDGNIRS, firehose \citep{2013PASP..125..270S}. ESOREFLEX \citep{esoreflex}, SciPy (\url{https://www.scipy.org/})}.

\appendix
\section{Photometric tables} \label{sec:phototables}

\begin{center}
\startlongtable
\begin{deluxetable*}{ccccccc}
\tablecaption{$BV$ and unfiltered light curves of DLT16am \label{table:BVphot}}
\tablehead{\colhead{Date} & \colhead{JD} & \colhead{phase} & \colhead{$B$(err)} & \colhead{$V$(err)} & \colhead{$Open$(err)} & \colhead{Inst} \\ 
\colhead{} & \colhead{} & \colhead{d} & \colhead{(mag)} & \colhead{(mag)} & \colhead{(mag)} & \colhead{} } 
\startdata
20161117 & 2457709.590 & -3.01  & $>19.8$ & $>20.6$     & \nodata     & IO:O        \\
20161117 & 2457710.070 & -2.53  & $>20.2$ & $>19.9$     & \nodata     & 1m0-03      \\
20161118 & 2457710.679 & -1.92  & \nodata & \nodata     & $>20.5$     & PROMPT5     \\
20161118 & 2457710.780 & -1.82  & \nodata & $>21.0$     & \nodata     & EFOSC2      \\
20161118 & 2457711.285 & -1.31  & $>20.8$ & $>20.5$     & \nodata     & 1m0-10      \\
20161119 & 2457711.510 & -1.09  & $>20.5$ & $>20.9$     & \nodata     & IO:O        \\
20161119 & 2457711.687 & -0.91  & \nodata & \nodata     & $>20.7$     & PROMPT5     \\
20161119 & 2457711.760 & -0.84  & \nodata & $>18.9$     & \nodata     & EFOSC2      \\
20161119 & 2457712.325 & -0.27  & $>20.5$ & $>20.0$     & \nodata     & 1m0-13      \\
20161120 & 2457712.515 & -0.08  & $>20.0$ & \nodata     & \nodata     & IO:O        \\
20161120 & 2457712.600 & 0.00   & \nodata & $>20.2$     & \nodata     & EFOSC       \\
20161121 & 2457713.530 & 0.93   & $>20.2$ & \nodata     & \nodata     & IO:O        \\
20161121 & 2457713.600 & 1.00   & \nodata & 20.84(0.52) & \nodata     & EFOSC       \\
20161121 & 2457713.686 & 1.09   & \nodata & \nodata     & 20.07(0.19) & PROMPT5     \\
20161121 & 2457714.220 & 1.62   & $>19.9$ & $>19.5$     & \nodata     & 1m0-03      \\
20161122 & 2457714.590 & 1.99   & \nodata & 19.72(0.10) & \nodata     & EFOSC2      \\
20161122 & 2457714.675 & 2.07   & \nodata & \nodata     & 18.57(0.06) & PROMPT5     \\
20161122 & 2457715.210 & 2.61   & \nodata & 19.34(0.13) & \nodata     & 1m0-11      \\
20161122 & 2457715.210 & 2.61   & $>20.5$ & \nodata     & \nodata     & 1m0-11      \\
20161123 & 2457715.540 & 2.94   & \nodata & \nodata     & 18.27(0.05) & PROMPT5     \\
20161123 & 2457716.045 & 3.44   & \nodata & 19.18(0.15) & \nodata     & 1m0-11      \\
20161123 & 2457716.045 & 3.44   & $>19.9$ & \nodata     & \nodata     & 1m0-11      \\
20161124 & 2457717.015 & 4.42   & \nodata & 19.24(0.13) & \nodata     & 1m0-03      \\
20161124 & 2457717.015 & 4.42   & $>20.4$ & \nodata     & \nodata     & 1m0-03      \\
20161125 & 2457717.530 & 4.93   & $>20.7$ & \nodata     & \nodata     & 1m0-05      \\
20161125 & 2457717.542 & 4.94   & \nodata & \nodata     & 18.14(0.05) & PROMPT5     \\
20161125 & 2457718.110 & 5.51   & $>19.7$ & \nodata     & \nodata     & 1m0-03      \\
20161125 & 2457718.420 & 5.82   & $>21.0$ & \nodata     & \nodata     & 1m0-10      \\
20161126 & 2457718.562 & 5.96   & \nodata & \nodata     & 18.02(0.05) & PROMPT5     \\
20161126 & 2457718.760 & 6.16   & \nodata & 19.27(0.22) & \nodata     & EFOSC2      \\
20161126 & 2457718.825 & 6.23   & \nodata & 19.21(0.11) & \nodata     & 1m0-04      \\
20161126 & 2457718.825 & 6.23   & $>20.2$ & \nodata     & \nodata     & 1m0-04      \\
20161127 & 2457719.543 & 6.94   & \nodata & \nodata     & 18.00(0.05) & PROMPT5     \\
20161127 & 2457720.290 & 7.69   & \nodata & 19.18(0.10) & \nodata     & 1m0-10      \\
20161127 & 2457720.290 & 7.69   & $>20.7$ & \nodata     & \nodata     & 1m0-10      \\
20161128 & 2457720.544 & 7.94   & \nodata & \nodata     & 17.91(0.05) & PROMPT5     \\
20161128 & 2457721.325 & 8.73   & \nodata & 19.19(0.11) & \nodata     & 1m0-13      \\
20161128 & 2457721.325 & 8.73   & $>20.7$ & \nodata     & \nodata     & 1m0-13      \\
20161129 & 2457721.535 & 8.94   & $>20.0$ & \nodata     & \nodata     & IO:O        \\
20161129 & 2457721.545 & 8.94   & \nodata & \nodata     & 17.68(0.05) & PROMPT5     \\
20161130 & 2457722.546 & 9.95   & \nodata & \nodata     & 17.69(0.05) & PROMPT5     \\
20161130 & 2457722.565 & 9.96   & \nodata & 19.08(0.09) & \nodata     & 1m0-10      \\
20161130 & 2457722.565 & 9.96   & $>21.1$ & \nodata     & \nodata     & 1m0-10      \\
20161130 & 2457723.415 & 10.81  & \nodata & 19.23(0.10) & \nodata     & 1m0-10      \\
20161130 & 2457723.415 & 10.81  & $>21.0$ & \nodata     & \nodata     & 1m0-10      \\
20161201 & 2457723.546 & 10.95  & \nodata & \nodata     & 17.92(0.04) & PROMPT5     \\
20161201 & 2457724.415 & 11.81  & \nodata & 19.20(0.12) & \nodata     & 1m0-12      \\
20161201 & 2457724.415 & 11.81  & $>20.7$ & \nodata     & \nodata     & 1m0-12      \\
20161202 & 2457724.546 & 11.95  & \nodata & \nodata     & 17.89(0.04) & PROMPT5     \\
20161202 & 2457725.310 & 12.71  & $>21.7$ & \nodata     & \nodata     & 1m0-10      \\
20161202 & 2457725.310 & 12.71  & \nodata & 19.25(0.10) & \nodata     & 1m0-10      \\
20161203 & 2457725.547 & 12.95  & \nodata & \nodata     & 18.01(0.04) & PROMPT5     \\
20161204 & 2457726.548 & 13.95  & \nodata & \nodata     & 18.00(0.04) & PROMPT5     \\
20161204 & 2457727.200 & 14.60  & $>18.1$ & $>19.0$     & \nodata     & 1m0-11      \\
20161205 & 2457728.470 & 15.87  & \nodata & 19.38(0.14) & \nodata     & IO:O        \\
20161205 & 2457728.470 & 15.87  & $>19.9$ & \nodata     & \nodata     & IO:O        \\
20161205 & 2457727.549 & 14.95  & \nodata & \nodata     & 17.90(0.07) & PROMPT5     \\
20161206 & 2457728.549 & 15.95  & \nodata & \nodata     & 17.99(0.05) & PROMPT5     \\
20161206 & 2457728.620 & 16.02  & \nodata & \nodata     & 18.02(0.04) & PROMPT5     \\
20161207 & 2457729.550 & 16.95  & \nodata & \nodata     & 18.10(0.04) & PROMPT5     \\
20161207 & 2457729.623 & 17.02  & \nodata & \nodata     & 18.03(0.04) & PROMPT5     \\
20161207 & 2457730.130 & 17.53  & \nodata & 19.42(0.13) & \nodata     & 1m0-03      \\
20161207 & 2457730.130 & 17.53  & $>20.8$ & \nodata     & \nodata     & 1m0-03      \\
20161207 & 2457730.475 & 17.88  & $>20.0$ & 19.36(0.13) & \nodata     & IO:O        \\
20161208 & 2457730.619 & 18.02  & \nodata & \nodata     & 17.46(0.04) & PROMPT5     \\
20161209 & 2457732.005 & 19.40  & \nodata & 19.49(0.12) & \nodata     & 1m0-03      \\
20161209 & 2457732.005 & 19.40  & $>20.6$ & \nodata     & \nodata     & 1m0-03      \\
20161210 & 2457732.530 & 19.93  & $>20.2$ & 19.42(0.12) & \nodata     & IO:O        \\
20161210 & 2457732.748 & 20.15  & \nodata & \nodata     & 18.27(0.06) & PROMPT5     \\
20161210 & 2457733.130 & 20.53  & $>20.3$ & \nodata     & \nodata     & 1m0-03      \\
20161210 & 2457733.370 & 20.77  & $>20.0$ & \nodata     & \nodata     & 1m0-10      \\
20161211 & 2457733.828 & 21.23  & \nodata & \nodata     & 18.12(0.04) & PROMPT5     \\
20161211 & 2457734.445 & 21.84  & \nodata & 19.44(0.14) & \nodata     & IO:O        \\
20161211 & 2457734.445 & 21.84  & $>19.8$ & \nodata     & \nodata     & IO:O        \\
20161212 & 2457734.553 & 21.95  & \nodata & \nodata     & 18.22(0.05) & PROMPT5     \\
20161213 & 2457735.500 & 22.90  & $>19.8$ & \nodata     & \nodata     & IO:O        \\
20161213 & 2457735.660 & 23.06  & \nodata & $>18.8$     & \nodata     & 1m0-05      \\
20161214 & 2457737.485 & 24.88  & $>19.3$ & $>19.2$     & \nodata     & 1m0-10      \\
20161215 & 2457738.480 & 25.88  & \nodata & 19.54(0.13) & \nodata     & 1m0-10      \\
20161216 & 2457738.625 & 26.02  & \nodata & \nodata     & 18.18(0.04) & PROMPT5     \\
20161217 & 2457739.556 & 26.96  & \nodata & \nodata     & 18.15(0.04) & PROMPT5     \\
20161218 & 2457740.556 & 27.96  & \nodata & \nodata     & 18.19(0.03) & PROMPT5     \\
20161218 & 2457740.730 & 28.13  & \nodata & 19.50(0.06) & \nodata     & EFOSC2      \\
20161219 & 2457741.556 & 28.96  & \nodata & \nodata     & 18.17(0.04) & PROMPT5     \\
20161220 & 2457742.500 & 29.90  & \nodata & 19.58(0.13) & \nodata     & 1m0-13      \\
20161220 & 2457742.556 & 29.96  & \nodata & \nodata     & 18.24(0.04) & PROMPT5     \\
20161220 & 2457743.450 & 30.85  & \nodata & 19.65(0.13) & \nodata     & 1m0-13      \\
20161221 & 2457744.370 & 31.77  & \nodata & 19.69(0.15) & \nodata     & 1m0-13      \\
20161222 & 2457744.557 & 31.96  & \nodata & \nodata     & 18.22(0.04) & PROMPT5     \\
20161223 & 2457745.686 & 33.09  & \nodata & \nodata     & 18.18(0.04) & PROMPT5     \\
20161224 & 2457746.557 & 33.96  & \nodata & \nodata     & 18.27(0.04) & PROMPT5     \\
20161225 & 2457747.575 & 34.98  & \nodata & \nodata     & 18.22(0.04) & PROMPT5     \\
20161225 & 2457748.470 & 35.87  & \nodata & $>19.5$     & \nodata     & 1m0-10      \\
20161226 & 2457749.450 & 36.85  & \nodata & 19.61(0.13) & \nodata     & 1m0-13      \\
20161227 & 2457749.558 & 36.96  & \nodata & \nodata     & 18.33(0.05) & PROMPT5     \\
20161227 & 2457750.350 & 37.75  & \nodata & 19.75(0.12) & \nodata     & 1m0-12      \\
20161228 & 2457750.558 & 37.96  & \nodata & \nodata     & 18.30(0.04) & PROMPT5     \\
20161228 & 2457751.450 & 38.85  & \nodata & 19.57(0.16) & \nodata     & 1m0-13      \\
20161229 & 2457751.559 & 38.96  & \nodata & \nodata     & 18.40(0.04) & PROMPT5     \\
20161229 & 2457752.350 & 39.75  & \nodata & 19.77(0.12) & \nodata     & 1m0-13      \\
20161230 & 2457752.559 & 39.96  & \nodata & \nodata     & 18.33(0.04) & PROMPT5     \\
20161230 & 2457753.495 & 40.90  & \nodata & 19.75(0.13) & \nodata     & 1m0-12      \\
20161230 & 2457753.495 & 40.90  & $>20.8$ & \nodata     & \nodata     & 1m0-12      \\
20161231 & 2457753.559 & 40.96  & \nodata & \nodata     & 18.28(0.04) & PROMPT5     \\
20170101 & 2457754.559 & 41.96  & \nodata & \nodata     & 18.35(0.04) & PROMPT5     \\
20170102 & 2457755.559 & 42.96  & \nodata & \nodata     & 18.44(0.05) & PROMPT5     \\
20170102 & 2457755.740 & 43.14  & $>21.1$ & \nodata     & \nodata     & 1m0-04      \\
20170103 & 2457756.562 & 43.96  & \nodata & \nodata     & 18.51(0.07) & PROMPT5     \\
20170103 & 2457756.715 & 44.11  & \nodata & 20.08(0.12) & \nodata     & 1m0-05      \\
20170103 & 2457756.715 & 44.11  & $>21.1$ & \nodata     & \nodata     & 1m0-05      \\
20170106 & 2457759.559 & 46.96  & \nodata & \nodata     & 18.46(0.06) & PROMPT5     \\
20170106 & 2457759.720 & 47.12  & \nodata & 19.75(0.06) & \nodata     & EFOSC2      \\
20170107 & 2457760.559 & 47.96  & \nodata & \nodata     & 18.42(0.08) & PROMPT5     \\
20170111 & 2457764.595 & 52.00  & \nodata & 19.93(0.15) & \nodata     & 1m0-05      \\
20170111 & 2457764.595 & 52.00  & $>21.3$ & \nodata     & \nodata     & 1m0-05      \\
20170116 & 2457769.671 & 57.07  & \nodata & \nodata     & 18.48(0.07) & PROMPT5     \\
20170117 & 2457770.558 & 57.96  & \nodata & \nodata     & 18.49(0.05) & PROMPT5     \\
20170118 & 2457771.558 & 58.96  & \nodata & \nodata     & 18.45(0.05) & PROMPT5     \\
20170118 & 2457771.590 & 58.99  & \nodata & 20.07(0.49) & \nodata     & EFOSC2      \\
20170119 & 2457772.557 & 59.96  & \nodata & \nodata     & 18.53(0.06) & PROMPT5     \\
20170119 & 2457772.635 & 60.03  & \nodata & 20.04(0.14) & \nodata     & 1m0-05      \\
20170119 & 2457772.635 & 60.03  & $>21.1$ & \nodata     & \nodata     & 1m0-05      \\
20170120 & 2457773.557 & 60.96  & \nodata & \nodata     & 18.49(0.05) & PROMPT5     \\
20170120 & 2457773.695 & 61.09  & \nodata & 20.21(0.15) & \nodata     & 1m0-05      \\
20170120 & 2457773.695 & 61.09  & $>22.4$ & \nodata     & \nodata     & 1m0-05      \\
20170121 & 2457774.685 & 62.08  & \nodata & 20.06(0.14) & \nodata     & 1m0-04      \\
20170121 & 2457774.685 & 62.08  & $>20.9$ & \nodata     & \nodata     & 1m0-04      \\
20170124 & 2457777.699 & 65.10  & \nodata & \nodata     & 18.28(0.08) & PROMPT5     \\
20170125 & 2457778.613 & 66.01  & \nodata & \nodata     & 18.71(0.07) & PROMPT5     \\
20170125 & 2457778.613 & 66.01  & \nodata & \nodata     & 19.15(0.08) & PROMPT5     \\
20170126 & 2457779.620 & 67.02  & \nodata & 20.22(0.33) & \nodata     & EFOSC2      \\
20170127 & 2457781.280 & 68.68  & \nodata & 20.03(0.18) & \nodata     & 1m0-13      \\
20170128 & 2457781.559 & 68.96  & \nodata & \nodata     & 18.57(0.07) & PROMPT5     \\
20170129 & 2457782.554 & 69.95  & \nodata & \nodata     & 18.48(0.06) & PROMPT5     \\
20170130 & 2457783.551 & 70.95  & \nodata & \nodata     & 18.58(0.06) & PROMPT5     \\
20170131 & 2457784.554 & 71.95  & \nodata & \nodata     & 18.67(0.07) & PROMPT5     \\
20170131 & 2457785.290 & 72.69  & \nodata & 20.19(0.16) & \nodata     & 1m0-13      \\
20170201 & 2457785.553 & 72.95  & \nodata & \nodata     & 18.73(0.07) & PROMPT5     \\
20170202 & 2457786.549 & 73.95  & \nodata & \nodata     & 18.65(0.06) & PROMPT5     \\
20170203 & 2457787.550 & 74.95  & \nodata & \nodata     & 18.60(0.07) & PROMPT5     \\
20170204 & 2457789.300 & 76.70  & \nodata & 20.16(0.16) & \nodata     & 1m0-13      \\
20170205 & 2457789.580 & 76.98  & \nodata & 20.36(0.34) & \nodata     & EFOSC2      \\
20170205 & 2457789.590 & 76.99  & \nodata & 20.14(0.20) & \nodata     & 1m0-04      \\
20170208 & 2457793.340 & 80.74  & \nodata & $>19.9$     & \nodata     & 1m0-13      \\
20170209 & 2457793.576 & 80.98  & \nodata & \nodata     & 18.81(0.07) & PROMPT5     \\
20170211 & 2457795.550 & 82.95  & \nodata & \nodata     & 18.82(0.06) & PROMPT5     \\
20170212 & 2457796.542 & 83.94  & \nodata & \nodata     & 19.06(0.09) & PROMPT5     \\
20170213 & 2457797.541 & 84.94  & \nodata & \nodata     & 19.12(0.11) & PROMPT5     \\
20170213 & 2457797.920 & 85.32  & \nodata & 20.58(0.18) & \nodata     & 1m0-03      \\
20170213 & 2457798.380 & 85.78  & \nodata & $>20.0$     & \nodata     & 1m0-10      \\
20170214 & 2457798.541 & 85.94  & \nodata & \nodata     & 19.17(0.08) & PROMPT5     \\
20170214 & 2457799.360 & 86.76  & \nodata & $>20.0$     & \nodata     & 1m0-13      \\
20170215 & 2457799.540 & 86.94  & \nodata & \nodata     & 19.54(0.11) & PROMPT5     \\
20170218 & 2457803.260 & 90.66  & \nodata & $>20.4$     & \nodata     & 1m0-13      \\
20170218 & 2457802.530 & 89.93  & \nodata & $>21.4$     & \nodata     & EFOSC2      \\
20170219 & 2457803.550 & 90.95  & \nodata & $>21.1$     & \nodata     & EFOSC2      \\
20170219 & 2457803.930 & 91.33  & \nodata & $>20.6$     & \nodata     & 1m0-03      \\
20170221 & 2457805.610 & 93.01  & \nodata & $>20.5$     & \nodata     & 1m0-05      \\
20170222 & 2457806.930 & 94.33  & \nodata & $>20.6$     & \nodata     & 1m0-03      \\
20170223 & 2457807.940 & 95.34  & \nodata & $>20.7$     & \nodata     & 1m0-11      \\
20170224 & 2457808.980 & 96.38  & \nodata & $>20.4$     & \nodata     & 1m0-11      \\
20170225 & 2457810.300 & 97.70  & \nodata & $>20.5$     & \nodata     & 1m0-13      \\
20170225 & 2457809.530 & 96.93  & \nodata & 20.99(0.56) & \nodata     & EFOSC       \\
20170226 & 2457810.900 & 98.30  & \nodata & $>20.5$     & \nodata     & 1m0-03      \\
20170301 & 2457814.300 & 101.70 & \nodata & $>20.9$     & \nodata     & 1m0-13      \\
20170302 & 2457815.300 & 102.70 & \nodata & $>20.5$     & \nodata     & 1m0-10      \\
20170303 & 2457816.290 & 103.69 & \nodata & $>20.6$     & \nodata     & 1m0-12      \\
20170306 & 2457818.500 & 105.90 & \nodata & $>21.0$     & \nodata     & EFOSC       \\
20170308 & 2457820.500 & 107.90 & \nodata & 21.14(0.06) & \nodata     & EFOSC       \\
20170312 & 2457824.920 & 112.32 & \nodata & $>19.8$     & \nodata     & 2m0-02      \\
\enddata
\tablecomments{
PROMPT5: $0.41\rm{m}$ PROMPT5 telescope at the Cerro Tololo Inter-American Observatory, Chile;
EFOSC2: $3.58\,\rm{m}$ ESO New Technology Telescope with EFOSC2 at the ESO La Silla Observatory, Chile;
IO:O: $2\,\rm{m}$ $2\,\rm{m}$ Liverpool Telescope with IO:O, at the Observatorio del Roque de Los Muchachos, Spain;
Las Cumbres Observatory 1m0-03, 1m0-11: node at Siding Spring, Australia; 1m0-04, 1m0-05, 2m0-02: node at Cerro Tololo Inter-American Observatory, Chile; 1m0-10, 1m0-12, 1m0-13: node at South African Astronomical Observatory, South Africa.}
\end{deluxetable*}

\startlongtable
\begin{deluxetable*}{cccccccc}
\tablecolumns{8}
\tablecaption{$griz$ light curves of DLT16am \label{table:grizphot}}
\tablehead{\colhead{Date} & \colhead{JD} & \colhead{phase} & \colhead{$g$(err)} & \colhead{$r$(err)} & \colhead{$i$(err)} & \colhead{$z$(err)} & \colhead{Inst} \\ 
\colhead{} & \colhead{} & \colhead{d} & \colhead{(mag)} & \colhead{(mag)} & \colhead{(mag)} & \colhead{(mag)} & \colhead{} } 
\startdata
20161116 & 2457708.60 & -4.00  & \nodata     & \nodata     & $>18.7$     & $>18.0$     & IO:O            \\
20161117 & 2457709.53 & -3.07  & \nodata     & $>20.4$     & $>19.9$     & $>19.8$     & GROND           \\
20161117 & 2457709.60 & -3.00  & \nodata     & \nodata     & $>19.3$     & $>18.4$     & IO:O            \\
20161118 & 2457710.53 & -2.08  & \nodata     & \nodata     & $>19.3$     & $>18.8$     & IO:O            \\
20161118 & 2457711.29 & -1.31  & $>21.7$     & \nodata     & \nodata     & \nodata     & 1m0-10          \\
20161118 & 2457711.30 & -1.31  & \nodata     & $>21.0$     & $>20.2$     & \nodata     & 1m0-10          \\
20161119 & 2457711.52 & -1.08  & \nodata     & \nodata     & $>19.4$     & $>18.7$     & IO:O            \\
20161119 & 2457711.78 & -0.82  & $>21.8$     & \nodata     & \nodata     & $>17.1$     & GROND           \\
20161119 & 2457712.34 & -0.27  & \nodata     & $>20.5$     & $>19.6$     & \nodata     & 1m0-13          \\
20161120 & 2457712.52 & -0.08  & \nodata     & \nodata     & $>19.0$     & $>18.6$     & IO:O            \\
20161121 & 2457713.53 & 0.92   & \nodata     & 20.37(0.45) & \nodata     & \nodata     & PROMPT1         \\
20161121 & 2457713.54 & 0.94   & \nodata     & \nodata     & $>18.9$     & $>18.4$     & IO:O            \\
20161121 & 2457713.61 & 1.01   & \nodata     & 20.18(0.23) & \nodata     & \nodata     & PROMPT1         \\
20161121 & 2457713.69 & 1.09   & \nodata     & 20.02(0.15) & \nodata     & \nodata     & PROMPT1         \\
20161121 & 2457713.81 & 1.21   & \nodata     & 19.88(0.22) & 18.95(0.16) & \nodata     & E2V           \\
20161121 & 2457714.23 & 1.63   & $>20.8$     & \nodata     & \nodata     & \nodata     & 1m0-03          \\
20161121 & 2457714.24 & 1.63   & \nodata     & 19.28(0.19) & 18.34(0.21) & \nodata     & 1m0-03          \\
20161122 & 2457714.55 & 1.94   & 20.87(0.28) & \nodata     & 17.94(0.17) & 17.17(0.10) & IO:O            \\
20161122 & 2457714.55 & 1.94   & \nodata     & 18.90(0.12) & \nodata     & \nodata     & PROMPT1         \\
20161122 & 2457714.63 & 2.03   & \nodata     & 18.89(0.10) & \nodata     & \nodata     & PROMPT1         \\
20161122 & 2457714.67 & 2.07   & 20.83(0.17) & 18.81(0.08) & 17.91(0.04) & 17.06(0.05) & GROND           \\
20161122 & 2457714.71 & 2.11   & \nodata     & 18.88(0.17) & \nodata     & \nodata     & PROMPT1         \\
20161122 & 2457714.80 & 2.20   & \nodata     & 18.83(0.11) & \nodata     & \nodata     & PROMPT1         \\
20161122 & 2457715.23 & 2.63   & \nodata     & 18.74(0.11) & 17.74(0.09) & \nodata     & 1m0-11          \\
20161123 & 2457715.63 & 3.03   & \nodata     & 18.76(0.11) & 17.60(0.08) & \nodata     & E2V           \\
20161123 & 2457715.73 & 3.13   & \nodata     & 18.73(0.12) & \nodata     & \nodata     & PROMPT1         \\
20161123 & 2457716.05 & 3.45   & 20.46(0.16) & \nodata     & \nodata     & \nodata     & 1m0-11          \\
20161123 & 2457716.06 & 3.46   & \nodata     & 18.46(0.07) & 17.51(0.08) & \nodata     & 1m0-11          \\
20161124 & 2457716.55 & 3.95   & \nodata     & 18.66(0.10) & \nodata     & \nodata     & PROMPT1         \\
20161124 & 2457716.60 & 4.00   & \nodata     & 18.69(0.25) & \nodata     & \nodata     & E2V           \\
20161124 & 2457716.68 & 4.08   & 20.65(0.10) & 18.61(0.07) & 17.27(0.05) & \nodata     & GROND           \\
20161124 & 2457717.02 & 4.42   & 20.47(0.14) & \nodata     & \nodata     & \nodata     & 1m0-03          \\
20161124 & 2457717.03 & 4.42   & \nodata     & 18.60(0.04) & 17.51(0.06) & \nodata     & 1m0-03          \\
20161125 & 2457717.54 & 4.94   & 20.52(0.14) & \nodata     & \nodata     & \nodata     & 1m0-05          \\
20161125 & 2457717.55 & 4.95   & \nodata     & 18.50(0.08) & 17.44(0.04) & \nodata     & 1m0-05          \\
20161125 & 2457717.69 & 5.09   & 20.44(0.04) & 18.61(0.07) & 17.43(0.03) & 16.53(0.04) & GROND           \\
20161125 & 2457717.76 & 5.16   & \nodata     & 18.54(0.11) & \nodata     & \nodata     & PROMPT1         \\
20161125 & 2457717.77 & 5.17   & \nodata     & 18.60(0.08) & 17.42(0.05) & \nodata     & E2V           \\
20161125 & 2457718.12 & 5.52   & 20.11(0.23) & \nodata     & \nodata     & \nodata     & 1m0-03          \\
20161125 & 2457718.12 & 5.52   & \nodata     & 18.48(0.08) & 17.43(0.06) & \nodata     & 1m0-03          \\
20161125 & 2457718.44 & 5.84   & 20.55(0.13) & \nodata     & \nodata     & \nodata     & 1m0-10          \\
20161125 & 2457718.45 & 5.85   & \nodata     & 18.47(0.06) & 17.38(0.03) & \nodata     & 1m0-10          \\
20161126 & 2457718.84 & 6.24   & 20.40(0.13) & \nodata     & \nodata     & \nodata     & 1m0-04          \\
20161126 & 2457718.85 & 6.25   & \nodata     & 18.51(0.07) & 17.33(0.06) & \nodata     & 1m0-04          \\
20161127 & 2457719.84 & 7.24   & \nodata     & 18.52(0.12) & \nodata     & \nodata     & PROMPT1         \\
20161127 & 2457720.31 & 7.71   & \nodata     & 18.51(0.06) & 17.31(0.04) & \nodata     & 1m0-10          \\
20161128 & 2457720.81 & 8.21   & \nodata     & 18.61(0.10) & 17.37(0.10) & \nodata     & E2V           \\
20161128 & 2457721.34 & 8.74   & 20.58(0.12) & \nodata     & \nodata     & \nodata     & 1m0-13          \\
20161128 & 2457721.35 & 8.75   & \nodata     & 18.47(0.12) & 17.27(0.05) & \nodata     & 1m0-13          \\
20161129 & 2457721.54 & 8.94   & 20.39(0.17) & 18.35(0.09) & 17.23(0.03) & 16.35(0.03) & IO:O            \\
20161129 & 2457721.62 & 9.02   & 20.55(0.08) & 18.45(0.05) & 17.24(0.01) & 16.32(0.02) & GROND           \\
20161129 & 2457721.72 & 9.11   & \nodata     & 18.60(0.15) & 17.33(0.14) & \nodata     & E2V           \\
20161130 & 2457722.58 & 9.98   & 20.49(0.13) & \nodata     & \nodata     & \nodata     & 1m0-10          \\
20161130 & 2457722.59 & 9.98   & \nodata     & 18.40(0.07) & 17.24(0.03) & \nodata     & 1m0-10          \\
20161130 & 2457722.71 & 10.11  & \nodata     & \nodata     & 17.16(0.20) & \nodata     & E2V           \\
20161130 & 2457723.43 & 10.83  & 20.49(0.10) & \nodata     & \nodata     & \nodata     & 1m0-10          \\
20161130 & 2457723.44 & 10.83  & \nodata     & 18.45(0.07) & 17.24(0.03) & \nodata     & 1m0-10          \\
20161130 & 2457723.50 & 10.90  & \nodata     & \nodata     & \nodata     & 16.28(0.11) & IO:O            \\
20161201 & 2457723.53 & 10.93  & \nodata     & 18.47(0.10) & \nodata     & \nodata     & PROMPT1         \\
20161201 & 2457723.81 & 11.21  & \nodata     & 18.45(0.10) & \nodata     & \nodata     & PROMPT1         \\
20161201 & 2457724.43 & 11.82  & 20.68(0.13) & \nodata     & \nodata     & \nodata     & 1m0-12          \\
20161201 & 2457724.44 & 11.83  & \nodata     & 18.45(0.05) & 17.22(0.03) & \nodata     & 1m0-12          \\
20161202 & 2457724.54 & 11.94  & \nodata     & 18.48(0.10) & \nodata     & \nodata     & PROMPT1         \\
20161202 & 2457724.76 & 12.16  & \nodata     & 18.50(0.10) & \nodata     & \nodata     & PROMPT1         \\
20161202 & 2457724.84 & 12.24  & \nodata     & 18.43(0.08) & \nodata     & \nodata     & PROMPT1         \\
20161202 & 2457725.32 & 12.72  & 20.45(0.12) & \nodata     & \nodata     & \nodata     & 1m0-10          \\
20161202 & 2457725.33 & 12.73  & \nodata     & 18.48(0.06) & 17.25(0.03) & 16.43(0.05) & 1m0-10          \\
20161204 & 2457726.73 & 14.13  & \nodata     & 18.25(0.17) & 17.49(0.16) & \nodata     & E2V           \\
20161204 & 2457727.21 & 14.61  & 20.40(0.24) & \nodata     & \nodata     & \nodata     & 1m0-11          \\
20161205 & 2457728.48 & 15.88  & 20.55(0.19) & 18.49(0.06) & 17.29(0.03) & 16.39(0.03) & IO:O            \\
20161206 & 2457728.53 & 15.93  & \nodata     & 18.55(0.10) & \nodata     & \nodata     & PROMPT1         \\
20161206 & 2457728.74 & 16.14  & \nodata     & 18.31(0.27) & \nodata     & \nodata     & E2V           \\
20161207 & 2457730.15 & 17.54  & 20.74(0.15) & \nodata     & \nodata     & \nodata     & 1m0-03          \\
20161207 & 2457730.15 & 17.55  & \nodata     & 18.62(0.11) & 17.41(0.05) & 16.62(0.09) & 1m0-03          \\
20161207 & 2457730.48 & 17.88  & $>19.8$     & \nodata     & \nodata     & \nodata     & IO:O            \\
20161207 & 2457730.48 & 17.88  & \nodata     & 18.54(0.13) & 17.35(0.05) & \nodata     & IO:O            \\
20161209 & 2457732.02 & 19.42  & 20.60(0.20) & \nodata     & \nodata     & \nodata     & 1m0-03          \\
20161209 & 2457732.03 & 19.43  & \nodata     & 18.58(0.09) & 17.47(0.04) & 16.77(0.05) & 1m0-03          \\
20161210 & 2457732.54 & 19.94  & 20.60(0.19) & 18.61(0.15) & 17.40(0.05) & 16.44(0.03) & IO:O            \\
20161210 & 2457733.15 & 20.55  & 20.65(0.25) & \nodata     & \nodata     & \nodata     & 1m0-03          \\
20161210 & 2457733.16 & 20.55  & \nodata     & 18.56(0.17) & 17.51(0.09) & 16.63(0.08) & 1m0-03          \\
20161210 & 2457733.41 & 20.80  & \nodata     & 18.70(0.14) & 17.56(0.09) & 16.63(0.11) & 1m0-10          \\
20161210 & 2457733.45 & 20.85  & \nodata     & \nodata     & 17.45(0.14) & 16.46(0.13) & IO:O            \\
20161211 & 2457734.45 & 21.84  & 20.63(0.17) & 18.64(0.12) & 17.49(0.07) & 16.49(0.03) & IO:O            \\
20161213 & 2457735.51 & 22.90  & $>19.7$     & \nodata     & \nodata     & \nodata     & IO:O            \\
20161213 & 2457735.51 & 22.90  & \nodata     & \nodata     & 17.50(0.08) & 16.50(0.08) & IO:O            \\
20161213 & 2457735.67 & 23.07  & 20.60(0.25) & \nodata     & \nodata     & \nodata     & 1m0-09          \\
20161213 & 2457736.47 & 23.86  & \nodata     & 18.55(0.10) & 17.48(0.04) & 16.50(0.03) & IO:O            \\
20161213 & 2457736.50 & 23.90  & \nodata     & 18.73(0.14) & 17.66(0.04) & 16.60(0.08) & ALFOSC\_FASU    \\
20161213 & 2457736.50 & 23.90  & $>19.8$     & \nodata     & \nodata     & \nodata     & ALFOSC\_FASU    \\
20161214 & 2457737.47 & 24.86  & \nodata     & \nodata     & 17.55(0.12) & \nodata     & IO:O            \\
20161215 & 2457737.50 & 24.90  & 20.61(0.41) & \nodata     & \nodata     & \nodata     & 1m0-10          \\
20161215 & 2457737.51 & 24.91  & \nodata     & 18.85(0.10) & 17.60(0.10) & 16.70(0.09) & 1m0-10          \\
20161215 & 2457738.50 & 25.90  & 20.99(0.22) & \nodata     & \nodata     & \nodata     & 1m0-10          \\
20161215 & 2457738.50 & 25.90  & \nodata     & 18.58(0.09) & 17.52(0.06) & 16.70(0.07) & 1m0-10          \\
20161220 & 2457743.46 & 30.86  & 21.30(0.18) & \nodata     & \nodata     & \nodata     & 1m0-13          \\
20161220 & 2457743.47 & 30.87  & \nodata     & 18.72(0.10) & 17.58(0.04) & 16.74(0.05) & 1m0-13          \\
20161220 & 2457742.73 & 30.13  & \nodata     & \nodata     & 17.54(0.07) & \nodata     & EFOSC           \\
20161221 & 2457744.38 & 31.78  & 21.15(0.18) & \nodata     & \nodata     & \nodata     & 1m0-13          \\
20161221 & 2457744.39 & 31.79  & \nodata     & 18.70(0.06) & 17.57(0.03) & 16.71(0.04) & 1m0-13          \\
20161223 & 2457745.72 & 33.11  & \nodata     & 18.58(0.30) & \nodata     & \nodata     & E2V           \\
20161225 & 2457748.48 & 35.88  & 21.28(0.30) & \nodata     & \nodata     & \nodata     & 1m0-10          \\
20161225 & 2457748.49 & 35.88  & \nodata     & 18.93(0.20) & 17.69(0.12) & 16.66(0.11) & 1m0-10          \\
20161226 & 2457749.46 & 36.86  & 21.02(0.17) & \nodata     & \nodata     & \nodata     & 1m0-13          \\
20161226 & 2457749.47 & 36.87  & \nodata     & 18.71(0.13) & 17.63(0.05) & 16.70(0.06) & 1m0-13          \\
20161227 & 2457750.36 & 37.76  & $>20.8$     & \nodata     & \nodata     & \nodata     & 1m0-12          \\
20161228 & 2457751.46 & 38.86  & 21.40(0.36) & \nodata     & \nodata     & \nodata     & 1m0-13          \\
20161228 & 2457751.47 & 38.87  & \nodata     & 18.57(0.14) & 17.64(0.08) & 16.73(0.10) & 1m0-13          \\
20161229 & 2457752.36 & 39.76  & 21.52(0.18) & \nodata     & \nodata     & \nodata     & 1m0-13          \\
20161229 & 2457752.37 & 39.77  & \nodata     & 18.76(0.07) & 17.64(0.03) & 16.72(0.04) & 1m0-13          \\
20161230 & 2457752.74 & 40.13  & \nodata     & 18.75(0.20) & \nodata     & \nodata     & E2V           \\
20161231 & 2457753.51 & 40.91  & 21.04(0.20) & \nodata     & \nodata     & \nodata     & 1m0-12          \\
20161231 & 2457753.52 & 40.92  & \nodata     & 18.75(0.10) & 17.70(0.06) & \nodata     & 1m0-12          \\
20170103 & 2457756.73 & 44.13  & $>21.1$     & \nodata     & \nodata     & \nodata     & 1m0-09          \\
20170103 & 2457756.74 & 44.13  & \nodata     & 18.76(0.06) & 17.63(0.08) & \nodata     & 1m0-05          \\
20170104 & 2457758.40 & 45.80  & \nodata     & \nodata     & 17.67(0.11) & \nodata     & IO:O            \\
20170111 & 2457764.62 & 52.02  & \nodata     & 18.78(0.06) & 17.70(0.04) & \nodata     & 1m0-05          \\
20170115 & 2457769.35 & 56.75  & \nodata     & 18.81(0.12) & 17.66(0.05) & \nodata     & IO:O            \\
20170116 & 2457770.41 & 57.80  & \nodata     & 18.87(0.05) & 17.75(0.01) & 16.71(0.01) & ALFOSC\_FASU    \\
20170116 & 2457770.41 & 57.80  & $>20.4$     & \nodata     & \nodata     & \nodata     & ALFOSC\_FASU    \\
20170119 & 2457772.66 & 60.05  & \nodata     & 18.82(0.08) & 17.76(0.03) & \nodata     & 1m0-05          \\
20170120 & 2457773.71 & 61.11  & $>21.0$     & \nodata     & \nodata     & \nodata     & 1m0-09          \\
20170120 & 2457773.72 & 61.11  & \nodata     & 18.87(0.09) & 17.73(0.05) & \nodata     & 1m0-05          \\
20170121 & 2457774.70 & 62.10  & 21.55(0.28) & \nodata     & \nodata     & \nodata     & 1m0-04          \\
20170121 & 2457774.71 & 62.10  & \nodata     & 18.93(0.09) & 17.76(0.04) & \nodata     & 1m0-04          \\
20170122 & 2457775.68 & 63.07  & \nodata     & 18.82(0.34) & \nodata     & \nodata     & E2V           \\
20170122 & 2457776.03 & 63.43  & 21.50(0.22) & \nodata     & \nodata     & \nodata     & 1m0-11          \\
20170122 & 2457776.04 & 63.44  & \nodata     & 18.95(0.09) & 17.80(0.05) & 16.81(0.06) & 1m0-11          \\
20170122 & 2457776.43 & 63.83  & 21.20(0.14) & \nodata     & \nodata     & \nodata     & 1m0-10          \\
20170122 & 2457776.45 & 63.84  & \nodata     & 18.84(0.08) & 17.75(0.04) & 16.86(0.06) & 1m0-10          \\
20170124 & 2457778.37 & 65.77  & $>20.4$     & $>17.6$     & \nodata     & \nodata     & IO:O            \\
20170126 & 2457779.62 & 67.02  & \nodata     & 18.92(0.08) & \nodata     & \nodata     & E2V           \\
20170126 & 2457780.34 & 67.73  & $>20.9$     & \nodata     & \nodata     & \nodata     & IO:O            \\
20170126 & 2457780.34 & 67.73  & \nodata     & 18.92(0.14) & 17.73(0.05) & \nodata     & IO:O            \\
20170127 & 2457780.65 & 68.04  & \nodata     & 18.96(0.08) & \nodata     & \nodata     & E2V           \\
20170127 & 2457781.29 & 68.69  & 21.45(0.24) & \nodata     & \nodata     & \nodata     & 1m0-13          \\
20170127 & 2457781.30 & 68.70  & \nodata     & 18.92(0.15) & 17.92(0.06) & 16.87(0.07) & 1m0-13          \\
20170128 & 2457782.34 & 69.74  & $>20.6$     & \nodata     & \nodata     & \nodata     & IO:O            \\
20170128 & 2457782.34 & 69.74  & \nodata     & 18.95(0.16) & 17.90(0.05) & \nodata     & IO:O            \\
20170131 & 2457785.31 & 72.71  & 21.69(0.20) & \nodata     & \nodata     & \nodata     & 1m0-13          \\
20170131 & 2457785.32 & 72.71  & \nodata     & 19.07(0.08) & 17.94(0.04) & 16.93(0.03) & 1m0-13          \\
20170202 & 2457787.38 & 74.78  & $>20.7$     & \nodata     & \nodata     & \nodata     & IO:O            \\
20170202 & 2457787.38 & 74.78  & \nodata     & 19.04(0.10) & 17.96(0.06) & \nodata     & IO:O            \\
20170204 & 2457789.31 & 76.71  & 22.05(0.47) & \nodata     & \nodata     & \nodata     & 1m0-13          \\
20170204 & 2457789.33 & 76.73  & \nodata     & 19.17(0.16) & 18.07(0.06) & 17.02(0.06) & 1m0-13          \\
20170205 & 2457789.60 & 77.00  & 22.01(0.29) & \nodata     & \nodata     & \nodata     & 1m0-04          \\
20170205 & 2457789.62 & 77.02  & \nodata     & 19.05(0.12) & 17.97(0.05) & 16.97(0.07) & 1m0-04          \\
20170207 & 2457792.39 & 79.79  & $>20.4$     & \nodata     & \nodata     & $>16.4$     & IO:O            \\
20170207 & 2457792.39 & 79.79  & \nodata     & 19.44(0.10) & 18.20(0.06) & \nodata     & IO:O            \\
20170208 & 2457792.66 & 80.05  & \nodata     & 18.92(0.32) & 18.07(0.10) & \nodata     & E2V           \\
20170208 & 2457793.36 & 80.76  & \nodata     & 19.25(0.13) & 18.13(0.13) & 17.14(0.09) & 1m0-13          \\
20170213 & 2457797.94 & 85.34  & $>21.4$     & \nodata     & \nodata     & \nodata     & 1m0-03          \\
20170213 & 2457797.95 & 85.35  & \nodata     & 19.56(0.18) & 18.53(0.13) & 17.33(0.15) & 1m0-03          \\
20170213 & 2457798.35 & 85.75  & $>21.2$     & \nodata     & \nodata     & \nodata     & 1m0-13          \\
20170213 & 2457798.36 & 85.75  & \nodata     & 19.62(0.10) & 18.52(0.10) & \nodata     & 1m0-13          \\
20170214 & 2457799.38 & 86.78  & $>21.5$     & \nodata     & \nodata     & \nodata     & 1m0-12          \\
20170214 & 2457799.39 & 86.79  & \nodata     & 20.00(0.14) & 18.76(0.14) & \nodata     & 1m0-12          \\
20170216 & 2457801.28 & 88.67  & \nodata     & 19.92(0.16) & 18.91(0.07) & \nodata     & 1m0-13          \\
20170218 & 2457802.63 & 90.02  & \nodata     & 20.33(0.37) & \nodata     & \nodata     & E2V           \\
20170218 & 2457803.29 & 90.69  & \nodata     & 20.24(0.11) & 19.22(0.14) & \nodata     & 1m0-13          \\
20170219 & 2457803.57 & 90.97  & $>21.6$     & \nodata     & \nodata     & \nodata     & 1m0-05          \\
20170220 & 2457804.57 & 91.97  & $>22.7$     & \nodata     & \nodata     & \nodata     & 1m0-09          \\
20170220 & 2457804.59 & 91.98  & \nodata     & 20.29(0.14) & 19.26(0.11) & \nodata     & 1m0-05          \\
20170220 & 2457804.63 & 92.03  & \nodata     & 20.37(0.29) & 19.18(0.16) & \nodata     & E2V           \\
20170221 & 2457805.92 & 93.32  & $>21.8$     & \nodata     & \nodata     & \nodata     & 1m0-03          \\
20170221 & 2457805.94 & 93.33  & \nodata     & 20.50(0.08) & 19.48(0.11) & \nodata     & 1m0-03          \\
20170222 & 2457806.59 & 93.98  & \nodata     & $>19.1$     & $>18.7$     & \nodata     & 1m0-08          \\  
20170223 & 2457807.63 & 95.02  & \nodata     & 20.67(0.08) & 19.51(0.10) & \nodata     & 1m0-05          \\
20170224 & 2457809.31 & 96.71  & $>21.9$     & \nodata     & \nodata     & \nodata     & 1m0-10          \\
20170224 & 2457809.32 & 96.71  & \nodata     & 20.73(0.15) & 19.55(0.19) & \nodata     & 1m0-10          \\
20170225 & 2457810.31 & 97.71  & $>21.6$     & \nodata     & \nodata     & \nodata     & 1m0-13          \\
20170225 & 2457810.32 & 97.72  & \nodata     & 20.69(0.10) & \nodata     & \nodata     & 1m0-13          \\
20170225 & 2457810.32 & 97.72  & \nodata     & \nodata     & $>19.5$     & \nodata     & 1m0-13          \\  
20170226 & 2457810.91 & 98.31  & $>21.9$     & \nodata     & \nodata     & \nodata     & 1m0-11          \\
20170226 & 2457810.93 & 98.32  & \nodata     & 20.74(0.10) & 19.59(0.15) & \nodata     & 1m0-11          \\
20170301 & 2457814.33 & 101.73 & \nodata     & 20.75(0.16) & 19.69(0.16) & 18.20(0.15) & 1m0-13          \\
20170302 & 2457815.32 & 102.72 & \nodata     & 20.74(0.09) & 19.71(0.11) & 18.20(0.12) & 1m0-10          \\
20170303 & 2457816.31 & 103.71 & \nodata     & 20.81(0.10) & 19.74(0.13) & 18.19(0.10) & 1m0-12          \\
20170305 & 2457817.56 & 104.96 & \nodata     & 20.83(0.55) & \nodata     & \nodata     & E2V           \\
20170306 & 2457818.56 & 105.96 & \nodata     & 20.92(0.08) & 19.74(0.14) & 18.28(0.16) & 1m0-05          \\
20170310 & 2457822.50 & 109.90 & \nodata     & 20.91(0.09) & \nodata     & \nodata     & E2V           \\
20170315 & 2457827.56 & 114.96 & \nodata     & $>19.54$    & \nodata     & \nodata     & E2V           \\
20170316 & 2457828.54 & 115.94 & \nodata     & $>19.79$    & \nodata     & \nodata     & E2V           \\
\enddata
\tablecomments{
IO:O: $2\,\rm{m}$ Liverpool Telescope with IO:O; GROND: MPG/ESO $2.2\,\rm{m}$ telescope with GROND at the ESO La Silla Observatory, Chile; PROMPT1: $0.41\,\rm{m}$ PROMPT1 telescope at the Cerro Tololo Inter-American Observatory, Chile; ALFOSC\_FASU: $2.56\,\rm{m}$ Nordic Optical Telescope with ALFOSC\_FASU, at the Observatorio del Roque de los Muchachos, Spain; E2V: Las Campanas Observatory $1\,\rm{m}$ Swope Telescope with the E2V camera, at the Las Campanas Observatory, Chile; Las Cumbres Observatory 1m0-03, 1m0-11: node at Siding Spring, Australia; 1m0-04, 1m0-05, 1m0-09: node at Cerro Tololo Inter-American Observatory, Chile; 1m0-10, 1m0-12, 1m0-13: node at South African Astronomical Observatory, South Africa.}
\end{deluxetable*}

\begin{deluxetable*}{ccccccc}
\tablecaption{NIR light curves of DLT16am \label{table:JHKphot}}
\tablehead{\colhead{JD} & \colhead{phase} & \colhead{$J$(err)} & \colhead{$H$(err)} & \colhead{$K$(err)} & \colhead{Inst} \\ 
           \colhead{}   & \colhead{}      & \colhead{(mag)}    & \colhead{(mag)}    & \colhead{(mag)}    & \colhead{} } 
\startdata
2457707.77 & -4.83  & $>18.3$	  & $>17.7$	& $>16.4$     & GROND   \\
2457708.73 & -3.87  & $>19.0$	  & $>18.2$	& $>16.7$     & GROND   \\
2457709.53 & -3.07  & $>18.4$	  & $>17.3$	& $>16.1$     & GROND   \\
2457711.78 & -0.82  & \nodata	  & $>15.9$	& \nodata     & GROND   \\
2457714.67 & 2.07   & 15.10(0.29) & 14.69(0.19) & 14.05(0.19) & GROND   \\
2457716.68 & 4.08   & 15.03(0.21) & 14.29(0.19) & 13.75(0.25) & GROND   \\
2457717.69 & 5.09   & 14.70(0.15) & 14.11(0.22) & 13.39(0.23) & GROND   \\
2457721.62 & 9.02   & 14.49(0.24) & 13.80(0.30) & 13.31(0.22) & GROND   \\
2457763.39 & 50.79  & 14.56(0.18) & 13.79(0.13) & 13.28(0.31) & NOTCam  \\
2457772.74 & 60.14  & 14.65(0.18) & 13.85(0.31) & 13.26(0.31) & SOFI    \\
2457790.68 & 78.07  & 14.85(0.21) & 13.90(0.32) & 13.47(0.21) & SOFI    \\
2457791.38 & 78.78  & 14.81(0.22) & 14.01(0.24) & 13.49(0.32) & NOTCam  \\
2457804.52 & 91.92  & 16.04(0.17) & 15.72(0.21) & 15.16(0.31) & SOFI    \\
2457819.52 & 106.92 & 16.40(0.15) & 15.61(0.27) & 14.79(0.30) & SOFI    \\
\enddata
\tablecomments{GROND: MPG/ESO $2.2\,\rm{m}$ telescope with GROND, at the ESO La Silla Observatory, Chile. NOTCam: $2.56\rm{m}$ Nordic Optical Telescope with NOTCam, at the Observatorio del Roque de los Muchachos, La Palma, Spain; SOFI: $3.58\,\rm{m}$ ESO New Technology Telescope with SOFI, at the ESO La Silla Observatory, Chile.}
\end{deluxetable*}
\end{center}

\end{document}